\documentclass{aa}
\usepackage{graphicx}
\usepackage{txfonts}
\usepackage{natbib}
\begin{document}
\title{Stability of the viscously spreading ring}
%   \subtitle{empty}
%%
\author{R. Speith\inst{1}
\and
W. Kley\inst{2}}
\offprints{R. Speith}
\institute{University of Leicester, Department of Physics and
Astronomy, Leicester LE1 7RH, United Kingdom\\
\email{ros@star.le.ac.uk}
\and
Universit\"at T\"ubingen, 
Institut f\"ur Astronomie und Astrophysik, 
Abt. Computational Physics,
Auf der Mor\-gen\-stelle 10, \mbox{D-72076} T\"ubingen, Germany\\
\email{kley@tat.physik.uni-tuebingen.de}
}
\date{Received 21 May 2002 / Accepted 2 December 2002}
\abstract{We study analytically and numerically the stability
of the pressure-less, viscously spreading accretion ring.
We show that the ring is unstable to small non-axisymmetric
perturbations. To perform the perturbation analysis of the ring
we use a stretching transformation of the time coordinate.
We find that to 1st order, one-armed spiral
structures, and to 2nd order additionally two-armed spiral features may
appear. Furthermore, we identify a dispersion relation determining the
instability of the ring. The theoretical results are confirmed in
several simulations, using two different numerical methods.
These computations prove independently the existence of a secular
spiral instability driven by viscosity, which evolves into
persisting leading and trailing spiral waves.
Our results settle the question whether the
spiral structures found in earlier simulations of the spreading ring
are numerical artifacts or genuine instabilities. 
}
\maketitle
%%-- Sect.1
\section{Introduction}

In the theory of accretion discs, the idealised problem of a
viscously spreading, pressure-less ring orbiting a central point mass
on Keplerian orbits is often used to exemplify the main features of
an evolving thin accretion disc, i.e. the inward transport of mass
and outward transport of angular momentum
\citep{1981ARA&A..19..137P,FrankKingRaine}. 
With the approximation of a small kinematic viscosity
$\nu$ that is independent of the surface density, 
an analytic solution exists for this problem, stated
originally by \citet{lues52} and later by \citet{1974MNRAS.168..603L}.

This axisymmetric analytic solution of the viscous dust ring is
frequently used as test problem for numerical methods developed to
simulate accretion discs. The tested algorithms cover particle methods
like smoothed particle hydrodynamics (SPH)
\citep[e.g.][]{1994ApJ...431..754F, Spei99} as well as grid-based
codes like {\it RH2D} \citep[e.g.][]{1999MNRAS.303..696K}.

However, numerical simulations of the evolving ring often show the
appearance of additional structures in the disc. In the majority of
cases, these structures consist of non-axisymmetric one-armed spirals,
but eventually also narrow concentric rings appear. Since their first
discovery, it is controversially debated whether these structures are
numerical artifacts or mathematical instabilities. While 
\citet{1996PASA...13...66M} 
show that for the axisymmetric case the concentric
rings found in SPH simulations are numerical artifacts of the method,
we demonstrate in this paper that the non-axisymmetric spiral structures
result from a genuine physical instability of the problem.

Recently, \citet{2001MNRAS.325..231O} analysed 
radially extended accretion discs
in a work presenting a general model of eccentric
accretion discs, of which the viscous ring is a simple special
case. However, because his main objective was much wider, he did not
elaborate on the stability properties of the ring.

The remainder of the paper is organised as follows. 
In the next section, we first summarise the basic hydrodynamic equations
used for our analysis of the evolving disc, and present for
reference the analytic solution of the viscously
spreading dust ring.

In Sect.~\ref{sec:perturbation} we use a special perturbation
technique, the time-stretching approach, which allows to obtain
solutions of the spreading ring at different orders of a small
expansion parameter. We derive a linear evolution equation for an
eccentricity function $E$. It is shown that to 2nd order, the general
solution allows the development of one-armed as well as two-armed
spiral features.

In Sect.~\ref{sec:local} we perform a local stability analysis of
the final equation for $E$ and deduce a dispersion relation which
indicates an instability for some viscosity laws. These results are
compared with numerical simulations we present in
Sect.~\ref{sec:numerics} 
using two different numerical methods, one
grid-based and the other particle based. 
We find in general very good agreement of both numerical methods
with the predictions of the local analysis.
Thus, the simulations confirm the existence of a physical instability
in the viscously evolving ring problem. 
Finally, in \ref{sec:conclusion} we give our conclusions.
\section{The equations}
\label{sec:equations}
In accretion physics one assumes that the evolution of the disc can be
modelled by the hydrodynamic equations. In the present case, we
envisage a thin accretion disc, i.e.\ the vertical thickness of the disc
is very small in comparison to the radial extend.
Therefore, the evolution of the surface density $\Sigma$ of
the disc can be modelled by the vertically averaged hydrodynamic
equations which reduces the problem to two dimensions and formally
corresponds to the 2D-version of the hydrodynamic equations.
\subsection{Basic hydrodynamic equations}
\label{sec:hydro}
Because of the axisymmetry of the ring solution, it is convenient to
work in polar coordinates $(R,\varphi)$. In the presence of a central
gravitational field originating by the mass $M_\mathrm{c}$, the set of the basic
hydrodynamic equations consists of the continuity equation 
\begin{equation}\label{ceq}
\frac{\partial\Sigma}{\partial t} + 
\frac{1}{R}\frac{\partial(R\Sigma v_R)}{\partial R} +
\frac{1}{R}\frac{\partial(\Sigma v_\varphi)}{\partial\varphi}
= 0
\quad, 
\end{equation}
the radial component of the Navier Stokes equation
\begin{eqnarray}
\lefteqn{
\Sigma\left(
\frac{\partial v_R}{\partial t} +
v_R\frac{\partial v_R}{\partial R} +
\frac{v_\varphi}{R}\frac{\partial v_R}{\partial\varphi} -
\frac{v_\varphi^2}{R}
\right)
= 
-\frac{\partial p_s}{\partial R} -
\Sigma\frac{G M_\mathrm{c}}{R^2}
}
\nonumber\\
& &
{}+
\frac{1}{R}\frac{\partial(R\nu_s\Sigma\sigma_{RR})}{\partial R} + 
\frac{1}{R}\frac{\partial(\nu_s\Sigma\sigma_{R\varphi})}{\partial\varphi} -
\frac{\nu_s\Sigma}{R}\sigma_{\varphi\varphi}
\label{ueq}
\quad,
\end{eqnarray} 
and the azimuthal component of the Navier Stokes equation
\begin{eqnarray}
\lefteqn{
\Sigma\left(
\frac{\partial v_\varphi}{\partial t} +
v_R\frac{\partial v_\varphi}{\partial R} +
\frac{v_\varphi}{R}\frac{\partial v_\varphi}{\partial\varphi} +
\frac{v_R v_\varphi}{R}
\right)
= -\frac{1}{R}\frac{\partial p_s}{\partial\varphi}
}
\nonumber\\
& &
{}+
\frac{1}{R}\frac{\partial(R\nu_s\Sigma\sigma_{R\varphi})}{\partial R} + 
\frac{1}{R}\frac{\partial(\nu_s\Sigma\sigma_{\varphi\varphi})}{\partial\varphi} +
\frac{\nu_s\Sigma}{R}\sigma_{R\varphi}
\label{weq}
\quad.
\end{eqnarray}
Here $v_R$ and $v_\varphi$ denote the radial and azimuthal velocity,
$p_s$ is the vertically averaged pressure, and $G$ is the
gravitational constant. Assuming that the bulk viscosity is
vanishing, the viscous shear tensor $\sigma$ has the form
\begin{eqnarray}
\sigma_{RR} & = & 
\frac{4}{3}\frac{\partial v_R}{\partial R} -
\frac{2}{3}\frac{v_R}{R} -
\frac{2}{3}\frac{1}{R}\frac{\partial v_\varphi}{\partial\varphi}
\label{shrr}\quad,\\
\sigma_{R\varphi} & = & \sigma_{\varphi R} \; = \;
\frac{\partial v_\varphi}{\partial R} -
\frac{v_\varphi}{R} +
\frac{1}{R}\frac{\partial v_R}{\partial\varphi}
\label{shrp}\quad,\\
\sigma_{\varphi\varphi} & = & 
-\frac{2}{3}\frac{\partial v_R}{\partial R} +
\frac{4}{3}\frac{v_R}{R} +
\frac{4}{3}\frac{1}{R}\frac{\partial v_\varphi}{\partial\varphi}
\label{shpp}\quad,
\end{eqnarray}
and $\nu_s$ is the vertically averaged kinematic viscosity. 

As we only consider cold (pressure-less) discs, we neglect the pressure
$p_s$ throughout the rest of this paper.

\subsection{Analytic solution of the viscous ring}

For the time evolution of the viscously spreading ring, we assume 
$\nu_s = \mathrm{const.}$ Then, with initial condition
\begin{equation}\label{initring}
\Sigma_\mathrm{init}(R) = \frac{M}{2\pi R_0}\delta(R-R_0)
\end{equation}
at $t = 0$ with total ring mass $M$, the surface density of the ring
evolves according to
\begin{equation}\label{asd}
\Sigma_\mathrm{ring}(\tau^*,x) = \frac{M}{\pi R_0^2}\frac{1}{\tau^*x^{1/4}}
\;I_{\frac{1}{4}}\!\!\left(\frac{2x}{\tau^*}\right)
\,\exp\!\left(-\frac{1+x^2}{\tau^*}\right)
\end{equation}
\citep[e.g.][]{1974MNRAS.168..603L},
where $x = R/R_0$, $\tau^* = 12\nu_s t/R_0^2$, and $I_{\frac{1}{4}}$
is the modified Bessel function to the order of 1/4.

The azimuthal velocity of the ring is equal to Keplerian velocity,
\begin{equation}\label{vkepler}
v_\mathrm{Kepler} = \sqrt{\frac{G M_\mathrm{c}}{R}} = \Omega R
\quad,
\end{equation}
where we have defined the angular velocity
\begin{equation}\label{angvel}
\Omega(R) = \sqrt{\frac{G M_\mathrm{c}}{R^3}}
\quad.
\end{equation}
The radial velocity of the ring obeys the relation
\begin{equation}\label{usol1}
v_\mathrm{spreading} = 
-\frac{3}{\Sigma_\mathrm{ring}\sqrt{R}}\frac{\partial}{\partial R}
\left[\nu_s\Sigma_\mathrm{ring}\sqrt{R}\right]
\quad.
\end{equation}

It is important to realize, that solution (\ref{asd}) fulfils the
hydrodynamic equations given in \ref{sec:hydro} only by approximation,
assuming 
the kinematic viscosity $\nu_s$ being small compared to the
specific angular momentum $\Omega R^2$. 

\section{Perturbation analysis}
\label{sec:perturbation}

To study the stability properties of the ring solution (\ref{asd}),
we perform a perturbation analysis of the hydrodynamic
equations (\ref{ceq}), (\ref{ueq}) and (\ref{weq}).

\subsection{Stretching transformation}

According to the idealised problem, we assume that the averaged
kinematic viscosity $\nu_s$ is very small and depends only on $R$.
To take this into account, we replace 
\begin{equation}\label{nusnu}
\nu_s = \nu_s(R) = \epsilon\,\nu(R)
\quad\mbox{where}\quad
\epsilon = \mathrm{const.} \ll 1
\quad.
\end{equation}
Then, two different timescales can be distinguished, the viscous
timescale $t_\mathrm{visc} \sim R^2 / \nu_s$ which determines the
radial spreading of the ring, and the dynamic timescale
$t_\mathrm{dyn} \sim R / v_\varphi = \sqrt{R^3 / G M_\mathrm{c}}$ describing
the azimuthal motion of the gas in the disc. Because of (\ref{nusnu}),
the dynamic timescale is much shorter than the viscous timescale, 
$t_\mathrm{dyn}/t_\mathrm{visc} \sim \nu_s / \sqrt{G M_\mathrm{c} R} = 
\epsilon\,\nu / (\Omega R^2) \ll 1$. Therefore, in addition to the
dynamic time $t$ we define a viscous time 
\begin{equation}
\tau = \epsilon\, t
\quad.
\end{equation}

In principle, the ring may evolve on the dynamic timescale as well as
on the viscous timescale. Thus we assume for the time dependencies of
all hydrodynamic quantities
\begin{equation}
\Sigma(t) = \hat\Sigma(t,\tau)
\,,\;
v_R(t) = \hat v_R(t,\tau)
\,,\;
v_\varphi(t) = \hat v_\varphi(t,\tau)
\;.
\end{equation}
This leads to a stretching transformation for the time derivatives,
\begin{equation}\label{strtra}
\frac{\partial f(t)}{\partial t} = 
\frac{\partial \hat f(t,\tau)}{\partial t} +
\epsilon\,\frac{\partial \hat f(t,\tau)}{\partial\tau}
\quad,
\end{equation}
where $f$ symbolises any of the hydrodynamic quantities.

\subsection{Approach}

To generalise the approach, we do not restrict the stability analysis
to a perturbation of the ring solution (\ref{asd}) but we start with
less determined unperturbed functions. The initial condition of the
problem shall consist of an arbitrary axisymmetric surface density
distribution $\Sigma_{\rm init}(R)$ rotating around a central mass
$M_\mathrm{c}$. Because self gravity of the disc can be neglected, the initial
azimuthal velocity shall be equal to the Keplerian velocity
(\ref{vkepler}), $v_\varphi^\mathrm{init} = v_\mathrm{Kepler}$. Then,
the following approach is made for the expansion
\begin{eqnarray}
\Sigma(t,R,\varphi) & = & 
\hat\Sigma_0(t,\tau,R) +
\epsilon\,\hat\Sigma_1(t,\tau,R,\varphi) 
\nonumber\\
& & {}+
\epsilon^2\,\hat\Sigma_2(t,\tau,R,\varphi)
+ {\cal O}(\epsilon^3)
\quad,
\label{sapp}\\
v_R(t,R,\varphi) & = & 
\hat v_{R0}(t,\tau,R) +
\epsilon\,\hat v_{R1}(t,\tau,R,\varphi) 
\nonumber\\
& & {}+
\epsilon^2\,\hat v_{R2}(t,\tau,R,\varphi)
+ {\cal O}(\epsilon^3)
\quad,
\label{uapp}\\
v_\varphi(t,R,\varphi) & = &
\hat v_{\varphi 0}(\tau,R) +
\epsilon\,\hat v_{\varphi 1}(t,\tau,R,\varphi) 
\nonumber\\
& & {}+
\epsilon^2\,\hat v_{\varphi 2}(t,\tau,R,\varphi)
+ {\cal O}(\epsilon^3)
\quad.
\label{wapp}
\end{eqnarray}
Here it is assumed that the 0th order quantities are axisymmetric and
do not depend on the azimuthal angle $\varphi$, and that the 0th order
azimuthal velocity evolves only very slowly and is independent of the
dynamical time $t$ (both assumptions are justified by the initial
conditions).

The approaches (\ref{sapp}), (\ref{uapp}) and (\ref{wapp}) are
inserted in (\ref{ceq}), (\ref{ueq}) and (\ref{weq}) considering the
stretching transformation (\ref{strtra}) and (\ref{nusnu}), and the
terms of the resulting equations are ordered according to powers of
$\epsilon$.

There are three main differences of the present approach
compared to previous analyses like those by
\citet{2001MNRAS.325..231O}. The kinematic viscosity is assumed to
be a function of radius only, while \citet{2001MNRAS.325..231O}
considered $\nu_s$ to be a function of surface density. Both dynamic
and viscous timescale are taken into account, while
\citet{2001MNRAS.325..231O} assumed at the outset that non of the
variables varies on the dynamic timescale. And the expansion is
carried out up to second order, so that non-linear terms appear in the
perturbation. Usually, a linear perturbation is performed. 

\subsection{Zeroth order results}

In ${\cal O}(1)$, the continuity equation reads
\begin{equation}\label{ceqe0}
\frac{\partial\hat\Sigma_0}{\partial t} + 
\frac{1}{R}\frac{\partial(R\hat\Sigma_0\hat v_{R0})}{\partial R}
= 0
\quad, 
\end{equation}
the radial component of the Navier Stokes equation reads
\begin{equation}\label{ueqe0}
\frac{\partial\hat v_{R0}}{\partial t} +
\hat v_{R0}\frac{\partial\hat v_{R0}}{\partial R} -
\frac{\hat v_{\varphi 0}^2}{R} 
= -\frac{G M_\mathrm{c}}{R^2}
\quad,
\end{equation}
and the azimuthal component accordingly
\begin{equation}\label{weqe0}
\frac{\hat v_{R0}}{R}\frac{\partial(R\hat v_{\varphi 0})}{\partial R}
= 0
\quad.
\end{equation}
From (\ref{weqe0}) follows immediately that the 0th order radial
velocity has to vanish everywhere,
\begin{equation}\label{u0}
\hat v_{R0} \equiv 0
\quad,
\end{equation}
because the alternative solution of (\ref{weqe0}), $\hat v_{\varphi 0}
\propto 1 / R$, does not fulfil the initial condition for the
azimuthal velocity. 

Inserting (\ref{u0}) into (\ref{ceqe0}) and (\ref{ueqe0}) results in 
\begin{equation}\label{dseqe0dt}
\frac{\partial\hat\Sigma_0}{\partial t} = 0
\quad,
\end{equation}
and
\begin{equation}\label{w0}
\hat v_{\varphi 0} = v_{\rm Kepler}
\quad,
\end{equation}
i.e., the 0th order of the azimuthal velocity is equal to the
Keplerian velocity, and the 0th order of the surface density does not
depend on the dynamical time, $\hat\Sigma_0 = \hat\Sigma_0(\tau,R)$.

\subsection{First order results}
\label{sec:pert1o}

In ${\cal O}(\epsilon)$, the continuity equation has the form
\begin{equation}\label{ceqe1}
\frac{\partial\hat\Sigma_1}{\partial t} + 
\Omega\frac{\partial\hat\Sigma_1}{\partial\varphi} + 
\frac{\partial\hat\Sigma_0}{\partial\tau} + 
\frac{1}{R}\frac{\partial(R\hat\Sigma_0\hat v_{R1})}{\partial R} +
\frac{\hat\Sigma_0}{R}\frac{\partial\hat v_{\varphi 1}}{\partial\varphi}
= 0
,
\end{equation}
and for the Navier Stokes equation results as radial component
\begin{equation}\label{ueqe1}
\frac{\partial\hat v_{R1}}{\partial t} + 
\Omega\left(\frac{\partial\hat v_{R1}}{\partial\varphi} - 
2\hat v_{\varphi 1} \right) = 0
\quad,
\end{equation}
and as azimuthal component
\begin{equation}\label{weqe1}
\frac{\partial\hat v_{\varphi 1}}{\partial t} + 
\frac{\Omega}{2}\left(2\frac{\partial\hat v_{\varphi 1}}{\partial\varphi}
+ \hat v_{R1} + 
\frac{3}{\hat\Sigma_0\sqrt{R}}\frac{\partial
\left[\nu\hat\Sigma_0\sqrt{R}\right]}{\partial R}
\right) = 0
.
\end{equation}

Taking the derivative of (\ref{weqe1}) with respect to $t$ considering
(\ref{dseqe0dt}) and solving for $\partial\hat v_{R1}/{\partial t}$,
and taking the derivative of (\ref{weqe1}) with respect to $\varphi$
and solving for $\partial\hat v_{R1}/{\partial\varphi}$, and inserting
the results into (\ref{ueqe1}) yields
\begin{equation}
\frac{\partial^2\hat v_{\varphi 1}}{\partial t^2} + 
2\Omega\frac{\partial^2\hat v_{\varphi 1}}{\partial t\partial\varphi}
+ \Omega^2\frac{\partial^2\hat v_{\varphi 1}}{\partial\varphi^2} 
+ \Omega^2\hat v_{\varphi 1} = 0
\quad.
\end{equation}
The general solution of this equation is
a linear superposition of
\begin{equation}\label{w1a}
\hat v_{\varphi 1} = \Re[v_{\varphi 1} \,\mathrm{e}^{\textstyle i(m\varphi - \omega t)}]
\end{equation}
with $v_{\varphi 1} = v_{\varphi 1}(\tau,R)$, 
and where $m$ and
$\omega$ have to obey the relation
\begin{equation}\label{relmom}
\omega^2 - 2m\Omega\omega + \Omega^2 m^2 - \Omega^2 = 0
\quad,
\end{equation}
that is
\begin{equation}
\omega = \Omega(m\pm 1)
\quad.
\end{equation}
Solving (\ref{weqe1}) for $\hat v_{R1}$ then results in
\begin{equation}\label{u1a}
\hat v_{R1} = \tilde v_{R1} + \Re[v_{R1} \,\mathrm{e}^{\textstyle i(m\varphi - \omega t)}]
\end{equation}
with
\begin{equation}\label{usol}
\tilde v_{R1} = - \frac{3}{\hat\Sigma_0\sqrt{R}}\frac{\partial}{\partial R}
\left[\nu\hat\Sigma_0\sqrt{R}\right]
\end{equation}
and
\begin{equation}
v_{R1} = 2i\left(\frac{\omega}{\Omega} - m\right) v_{\varphi 1}
\quad.
\end{equation}

Inserting (\ref{w1a}) and (\ref{u1a}) into (\ref{ceqe1}) gives
\begin{eqnarray}
\lefteqn{
\frac{\partial\hat\Sigma_1}{\partial t} + 
\Omega\frac{\partial\hat\Sigma_1}{\partial\varphi} 
=
{}-\left[\frac{\partial\hat\Sigma_0}{\partial\tau} + 
\frac{1}{R}\frac{\partial(R\hat\Sigma_0\tilde v_{R1})}{\partial R}
\right]
}
\label{ceqe1a}\\
& &
{}- \Re[
i\hat\Sigma_0 v_{R1}\left(\varphi\frac{\partial m}{\partial R} 
- t\frac{\partial\omega}{\partial R}\right)\,\mathrm{e}^{\textstyle
i(m\varphi - \omega t)}
]
\nonumber
\\
& &
{}-\Re[\left[\frac{1}{R}\frac{\partial(R\hat\Sigma_0 v_{R1})}{\partial R} +
im\frac{\hat\Sigma_0}{R} v_{\varphi 1}\right]\,\mathrm{e}^{\textstyle i(m\varphi - \omega
t)}]
\nonumber
\end{eqnarray}
The first term in brackets on the right hand side of (\ref{ceqe1a})
and the terms in parenthesis in the second line of the right hand side
of (\ref{ceqe1a}) are secular terms for $\hat\Sigma_1$. They have to
vanish identically, otherwise $\hat\Sigma_1$ would increase linearly
or, for the latter, even quadratically with $t$ or $\varphi$,
respectively.

Then, (\ref{ceqe1a}) can be solved for $\hat\Sigma_1$, 
\begin{equation}\label{s1a}
\hat\Sigma_1 = \Re[\Sigma_1 \,\mathrm{e}^{\textstyle i(m\varphi - \omega
t)}]
\end{equation}
with
\begin{equation}
\Sigma_1 = \frac{1}{R(\Omega m - \omega)}
\left[i\frac{\partial(R\hat\Sigma_0 v_{R1})}{\partial R} 
- m\hat\Sigma_0 v_{\varphi 1}\right]
\quad.
\end{equation}

The conditions for the disappearance of the secular terms in
parenthesis in (\ref{ceqe1a}) are
\begin{eqnarray}
\frac{\partial m}{\partial R} & = & 0 
\qquad\mbox{and}\\
\frac{\partial\omega}{\partial R} & = & 0
\quad.
\end{eqnarray}
Because of relation (\ref{relmom}) and $\Omega = \Omega(R)$, this can
only be achieved by
\begin{equation}\label{omeq0}
\omega \equiv 0
\quad,
\end{equation}
which leads to $m^2 = 1$.

The condition for the disappearance of the secular term in brackets in
(\ref{ceqe1a}) is
\begin{eqnarray}\label{ssol}
\frac{\partial\hat\Sigma_0}{\partial\tau} 
& = &
-\frac{1}{R}\frac{\partial(R\hat\Sigma_0\tilde v_{R1})}{\partial R} 
\\
& = &
\frac{3}{R}\frac{\partial}{\partial R}\left[
\sqrt{R}\frac{\partial}{\partial R}\left(\nu\hat\Sigma_0\sqrt{R}
\right)\right]
\quad.
\nonumber
\end{eqnarray}
This is the diffusion equation of the surface density for the
analytic solution of the viscous dust ring. With initial condition 
(\ref{initring}) and constant viscosity $\nu$, the surface density
(\ref{asd}) solves (\ref{ssol}) exactly (considering $\tau^* =
12\nu\tau/R_0^2$). 

That is, the surface density of the analytic solution is equal to the
0th order term of the perturbation. The azimuthal velocity of the
analytic solution is also equal to the according 0th order term
of the perturbation, i.e., equal to the Keplerian velocity (\ref{w0}).
And the radial velocity of the analytic solution is equal to the
axisymmetric part of the according 1st order term of the perturbation,
i.e., relations (\ref{usol1}) and (\ref{usol}) are identical.

Because of (\ref{omeq0}), all first order quantities are independent
of the dynamic time $t$. Therefore, the first order velocities take
the form
\begin{eqnarray}
\hat v_{R1} & = & \tilde v_{R1} + \Re[v_{R1} \,\mathrm{e}^{\textstyle i m\varphi}]
\qquad\mbox{and}\label{u1}\\
\hat v_{\varphi 1} & = & \Re[v_{\varphi 1} \,\mathrm{e}^{\textstyle i m\varphi}]
\; = \; \Re[\frac{i}{2}m v_{R1} \,\mathrm{e}^{\textstyle i m\varphi}]
\label{w1}\\
\mbox{with}\qquad m & = & \pm 1
\label{meq1}
\quad.
\end{eqnarray}
Thus, if a non-axisymmetric perturbation appears, to first order it
has to be an one-armed spiral structure. Additionally, it follows
\begin{equation}\label{factor2}
|v_{R1}| = 2|v_{\varphi 1}|
\quad,
\end{equation}
and $v_{R1}$ and $v_{\varphi 1}$ obey a phase shift of
$\frac{\pi}{2}$.

Defining a function $\hat E = \hat E(\tau,R)$ similar to 
\citet{2001MNRAS.325..231O}
with
\begin{equation}\label{uef}
v_{R1}(\tau,R) = i m R\Omega(R) \hat E(\tau,R)
\end{equation}
yields
\begin{equation}\label{wef}
v_{\varphi 1}(\tau,R) = -\frac{1}{2}R\Omega(R) \hat E(\tau,R)
\quad.
\end{equation}
Then, with (\ref{s1a}), the first order surface density takes the form
\begin{equation}\label{s1}
\hat\Sigma_1(\tau,R,\varphi) = \Re[\Sigma_1(\tau,R)\,\mathrm{e}^{\textstyle i m\varphi}]
\quad,
\end{equation}
where
\begin{equation}\label{s1sol}
\Sigma_1 = 
\frac{i}{R\Omega m}\left[
\frac{\partial(R\hat\Sigma_0 v_{R1})}{\partial R} - 
\frac{\hat\Sigma_0 v_{R1}}{2}\right] =
- R\frac{\partial(\hat\Sigma_0 \hat E)}{\partial R}
\;.
\end{equation}

Furthermore, an initial phase $\varphi_0$ can be found such that $\hat
E(\tau,R) = E(\tau,R)\,\mathrm{e}^{\textstyle i\varphi_0}$ with $E =
|\hat E|$.  

\subsection{Second order results}
\label{sec:pert2o}

Consider the 0th order results (\ref{u0}) and (\ref{w0}) and the
1st order results (\ref{u1}), (\ref{w1}) and (\ref{s1}). Then, in 
${\cal O}(\epsilon^2)$ one yields for the continuity equation
\begin{eqnarray}\label{ceqe2}
\lefteqn{
\frac{\partial\hat\Sigma_2}{\partial t} + 
\Omega\frac{\partial\hat\Sigma_2}{\partial\varphi} + 
\frac{1}{R}\frac{\partial(R\hat\Sigma_0\hat v_{R2})}{\partial R} +
\frac{\hat\Sigma_0}{R}\frac{\partial\hat v_{\varphi
2}}{\partial\varphi}
}
\\
& = &
{}-\frac{m}{2R}\frac{\partial}{\partial R}\left[
R^2\Omega E
\frac{\partial(\hat\Sigma_0 E)}{\partial R}
\right]\,\sin(2m\varphi+2\varphi_0)
\nonumber\\
& &
{}+ \left(
R\frac{\partial}{\partial R}\left[
\hat\Sigma_0\frac{\partial E}{\partial\tau}
\right] 
- 
R\frac{\partial}{\partial R}\left[
\frac{E}{R}
\frac{\partial(R\hat\Sigma_0\tilde v_{R1})}{\partial R}
\right]
\right.
\nonumber\\
& &
\left.
{} + 
\frac{1}{R}\frac{\partial}{\partial R}\left[
R^2\tilde v_{R1}
\frac{\partial(\hat\Sigma_0 E)}{\partial R}
\right]
\right)\,\cos(m\varphi+\varphi_0)
\quad,
\nonumber
\end{eqnarray}
for the radial component of the Navier Stokes equation
\begin{eqnarray}
\lefteqn{
\hat\Sigma_0\left[
\frac{\partial\hat v_{R2}}{\partial t} + 
\Omega\left(\frac{\partial\hat v_{R2}}{\partial\varphi} - 
2\hat v_{\varphi 2} \right) 
\right]} 
\label{ueqe2}\\
\lefteqn{
{}-m\hat\Sigma_0 R\Omega\frac{\partial
E}{\partial\tau}\,\sin(m\varphi+\varphi_0)
}
\nonumber\\
\lefteqn{
{}+\frac{1}{2}\hat\Sigma_0 R^2\Omega^2 E\left[\left(
\frac{3}{4}\frac{E}{R} - \frac{\partial E}{\partial R}
\right)\cos(2m\varphi+2\varphi_0)
\right.
}
\nonumber\\
\lefteqn{\qquad
\left.
{}- \frac{1}{4}\frac{E}{R} + \frac{\partial E}{\partial R}\right]
}
\nonumber\\
\lefteqn{
{}-m\Omega\left[
\frac{E}{R}\left(
\frac{1}{R^2}\frac{\partial(R^3\nu\hat\Sigma_0)}{\partial R}
\right.\right.
}
\nonumber\\
\lefteqn{\qquad
\left.
{}+
3R^2\left[
\frac{1}{\hat\Sigma_0}\frac{\partial\hat\Sigma_0}{\partial R}
\frac{\partial(\nu\hat\Sigma_0)}{\partial R}
-
\frac{\partial^2(\nu\hat\Sigma_0)}{\partial R^2}
\right]
\right)
}
\nonumber\\
\lefteqn{\qquad
\left.
{}-\frac{1}{3}\frac{\partial E}{\partial R}\left(
13R\frac{\partial(\nu\hat\Sigma_0)}{\partial R}
+
\frac{17}{2}\nu\hat\Sigma_0
\right)
\right.
}
\nonumber\\
\lefteqn{\qquad
{}-
\frac{4}{3}\nu\hat\Sigma_0 R\frac{\partial^2 E}{\partial R^2}
\Bigg]\sin(m\varphi+\varphi_0)
}
\nonumber\\
& = & 
\frac{4}{3}\frac{1}{R}
\frac{\partial}{\partial R}\left[
\nu\hat\Sigma_0 R\sqrt{R}\frac{\partial}{\partial R}
\left(\frac{\tilde v_{R1}}{\sqrt{R}}\right)
\right]
\nonumber\\
& &
{}+ 
\frac{2}{3}\nu\hat\Sigma_0 R
\frac{\partial}{\partial R}
\left(\frac{\tilde v_{R1}}{R^2}\right)
-
\hat\Sigma_0\left(
\frac{\partial\tilde v_{R1}}{\partial\tau}+
\tilde v_{R1}\frac{\partial\tilde v_{R1}}{\partial R}
\right)
\nonumber
\end{eqnarray}
where for some transformations the relation (\ref{usol}) is used, and
for the azimuthal component of the Navier Stokes equation
\begin{eqnarray}
\lefteqn{
\hat\Sigma_0\left[
\frac{\partial\hat v_{\varphi 2}}{\partial t} + 
\Omega\left(\frac{\partial\hat v_{\varphi 2}}{\partial\varphi} +
\frac{1}{2}\hat v_{R2} \right) 
\right] 
}
\label{weqe2}\\
\lefteqn{
{}-\frac{1}{2}\hat\Sigma_0 R\Omega
\frac{\partial E}{\partial\tau}\,\cos(m\varphi+\varphi_0)
}
\nonumber\\
\lefteqn{
{}+\frac{1}{4}m\hat\Sigma_0 R^2\Omega^2 E
\frac{\partial E}{\partial R}\,\sin(2m\varphi+2\varphi_0)
}
\nonumber\\
& = &
\Omega\left[
\frac{E}{R}\left(
{}-\frac{3}{2}\nu\hat\Sigma_0 +
\frac{1}{2}R\nu\frac{\partial\hat\Sigma_0}{\partial R} -
R\hat\Sigma_0\frac{\partial\nu}{\partial R}
\right.\right.
\nonumber\\
& &
\qquad
{}-
\frac{3}{2}R^2\nu
\left[
\frac{\partial\hat\Sigma_0}{\partial R}
\right]^2
+
\frac{3}{2}R^2
\nu\frac{\partial^2\hat\Sigma_0}{\partial R^2}
\Bigg)
\nonumber\\
& & 
{}+\frac{\partial E}{\partial R}\left(
\frac{5}{12}\nu\hat\Sigma_0 -
\frac{1}{2}R\nu\frac{\partial\hat\Sigma_0}{\partial R} -
2 R\hat\Sigma_0\frac{\partial\nu}{\partial R}
\right)
\nonumber\\
& &
{}+
\nu\hat\Sigma_0 R\frac{\partial^2 E}{\partial R^2}
\Bigg]\cos(m\varphi+\varphi_0)
\quad,
\nonumber
\end{eqnarray}
again considering relation (\ref{usol}).

Solving (\ref{ueqe2}) for $\hat v_{\varphi 2}$ and inserting in
(\ref{weqe2}) yields
\begin{eqnarray}\label{w2aeq}
\frac{\hat\Sigma_0}{2\Omega}\left[
\frac{\partial^2\hat v_{R2}}{\partial t^2} + 
2\Omega\frac{\partial^2\hat v_{R2}}{\partial t\partial\varphi}
+ \Omega^2\left(
\frac{\partial^2\hat v_{R2}}{\partial\varphi^2} 
+ \hat v_{R2}
\right)\right] 
\\
{}+ \frac{1}{2}\hat\Sigma_0\Omega \big(
U_1\cos(m\varphi+\varphi_0) + U_2\sin(2m\varphi+2\varphi_0) 
\big)
& = & 0
\nonumber
\end{eqnarray}
with
\begin{eqnarray}
\lefteqn{
U_1(\tau,R)
}
\label{bu1}\\
& = &
\frac{E}{R}\left(
R\frac{\partial\nu}{\partial R} + 
3R^2\frac{\partial^2\nu}{\partial R^2} -
2R\frac{\nu}{\hat\Sigma_0}\frac{\partial\hat\Sigma_0}{\partial R} +
3\frac{R^2}{\hat\Sigma_0}
\frac{\partial\nu}{\partial R}
\frac{\partial\hat\Sigma_0}{\partial R}
\right)
\nonumber\\
& &
{} + \frac{\partial E}{\partial R}\left(
2\nu + 
\frac{25}{3}R\frac{\partial\nu}{\partial R} +
\frac{16}{3}R\frac{\nu}{\hat\Sigma_0}
\frac{\partial\hat\Sigma_0}{\partial R}
\right) 
\nonumber\\
& &
{}-
\frac{2}{3}\nu R\frac{\partial^2 E}{\partial R^2}
-
2R\frac{\partial E}{\partial\tau}
\nonumber
\end{eqnarray}
and
\begin{equation}\label{bu2}
U_2(\tau,R) = 
\frac{3}{4}m R^2\Omega E
\left(2\frac{\partial E}{\partial R} - \frac{E}{R}\right)
\quad.
\end{equation}
A solution of (\ref{w2aeq}) is
\begin{eqnarray}
\hat v_{R2} & = & \Re[v_{R2}\,\mathrm{e}^{\textstyle i(\tilde m\varphi -
\tilde\omega t)}]
\label{u2a}\\
& &
{}- \frac{1}{2}U_1\varphi \sin(m\varphi+\varphi_0)
+ \frac{1}{3}U_2\sin(2m\varphi+2\varphi_0)
\nonumber
\end{eqnarray}
with $v_{R2} = v_{R2}(\tau,R)$, 
and with
\begin{equation}\label{reltmom}
\tilde\omega = \Omega(\tilde m\pm 1)
\quad.
\end{equation}
Inserting (\ref{u2a}) into (\ref{ueqe2}) results in 
\begin{eqnarray}\label{w2a}
\hat v_{\varphi 2} & = & \tilde v_{\varphi 2} + 
\Re[
v_{\varphi 2}\,\mathrm{e}^{\textstyle i(\tilde m\varphi - \tilde\omega
t)}]
\\
& &
{}+ \frac{1}{4}W_1\sin(m\varphi+\varphi_0) - 
\frac{1}{4}m U_1\varphi\cos(m\varphi+\varphi_0)
\nonumber\\
& &
{}+ W_2\,\big(\cos(2m\varphi+2\varphi_0) + 1\big)
\nonumber
\end{eqnarray}
with
\begin{eqnarray}
\tilde v_{\varphi 2} & = &
\frac{1}{2\Omega}\left(
\frac{\partial\tilde v_{R1}}{\partial\tau}+
\tilde v_{R1}\frac{\partial\tilde v_{R1}}{\partial R}
\right)
- 
\frac{R\nu}{3\Omega}
\frac{\partial}{\partial R}
\left(\frac{\tilde v_{R1}}{R^2}\right)
\\
& &
{}-
\frac{2}{3 R\hat\Sigma_0\Omega}
\frac{\partial}{\partial R}\left[
\nu\hat\Sigma_0 R\sqrt{R}\frac{\partial}{\partial R}
\left(\frac{\tilde v_{R1}}{\sqrt{R}}\right)
\right]
\quad,
\nonumber
\end{eqnarray}
\begin{equation}
v_{\varphi 2} = 
-\frac{1}{2}i\left(\frac{\tilde\omega}{\Omega} - \tilde m\right)
v_{R2}
\quad,
\end{equation}
\begin{eqnarray}
\lefteqn{
W_1(\tau,R)
}
\label{bw1}\\
& = &
\frac{E}{R}\Bigg(
-6m\nu - 
(2m + 1)R\frac{\partial\nu}{\partial R} + 
(6m - 3)R^2\frac{\partial^2\nu}{\partial R^2}
\nonumber\\
& &
{} -
2(m - 1)R\frac{\nu}{\hat\Sigma_0}\frac{\partial\hat\Sigma_0}{\partial
R} +
(6m - 3)\frac{R^2}{\hat\Sigma_0}
\frac{\partial\nu}{\partial R}
\frac{\partial\hat\Sigma_0}{\partial R} 
\nonumber\\
& &
{} -
6mR^2\frac{\nu}{\hat\Sigma_0^2}
\left[
\frac{\partial\hat\Sigma_0}{\partial R}
\right]^2 +
6mR^2\frac{\nu}{\hat\Sigma_0}
\frac{\partial^2\hat\Sigma_0}{\partial R^2}
\Bigg)
\nonumber\\
& &
{} + \frac{\partial E}{\partial R}\Bigg(%\left(
\frac{17m - 6}{3}\nu + 
\frac{26m - 25}{3}R\frac{\partial\nu}{\partial R} 
\nonumber\\
& & \left.
{}
+
\frac{26m - 16}{3}R\frac{\nu}{\hat\Sigma_0}
\frac{\partial\hat\Sigma_0}{\partial R}
\right) 
\nonumber\\
& &
{}+\left(\frac{2}{3} + \frac{8m}{3}\right)
\nu R\frac{\partial^2 E}{\partial R^2}
-
2(m-1)R\frac{\partial E}{\partial\tau}
\quad,
\nonumber
\end{eqnarray}
and
\begin{equation}
W_2(\tau,R) = 
\frac{1}{4}R^2\Omega E
\left(3\frac{\partial E}{\partial R} - \frac{5}{4}\frac{E}{R}\right)
\quad.
\end{equation}

Inserting (\ref{u2a}) and (\ref{w2a}) into the continuity equation
(\ref{ceqe2}) again results in a secular term of the form
\begin{equation}
\propto
\left(\varphi\frac{\partial\tilde m}{\partial R} 
- t\frac{\partial\tilde\omega}{\partial R}\right)
\quad,
\end{equation}
which has to vanish. Like in the 1st order case, this can only be
achieved by
\begin{equation}\label{tomvan}
\tilde\omega \equiv 0
\quad.
\end{equation}
Again this leads to $\tilde m^2 = 1$. Hence follows that all
quantities in 2nd order, $\hat v_{\varphi 2}$, $\hat v_{R2}$ and
$\hat\Sigma_2$, are also independent of the dynamical time $t$, and
$\hat\Sigma_2$ also can be integrated directly.

There exists a second secular term in (\ref{ceqe2}), which is
appearing already in (\ref{u2a}) and (\ref{w2a}) and which is
proportional to $U_1\varphi$. Therefore, $U_1$ has to vanish,
\begin{equation}
U_1 \equiv 0
\quad.
\end{equation}
This leads to a differential equation for $E(\tau,R)$,
\begin{eqnarray}
\frac{\partial E}{\partial\tau}
& = &
\frac{E}{R}\left(
\frac{1}{2}\frac{\partial\nu}{\partial R} + 
\frac{3}{2}R\frac{\partial^2\nu}{\partial R^2} -
\frac{\nu}{\hat\Sigma_0}\frac{\partial\hat\Sigma_0}{\partial R} +
\frac{3}{2}\frac{R}{\hat\Sigma_0}
\frac{\partial\nu}{\partial R}
\frac{\partial\hat\Sigma_0}{\partial R}
\right)
\nonumber\\
& &
{} + \frac{\partial E}{\partial R}\left(
\frac{\nu}{R} + 
\frac{25}{6}\frac{\partial\nu}{\partial R} +
\frac{8}{3}\frac{\nu}{\hat\Sigma_0}
\frac{\partial\hat\Sigma_0}{\partial R}
\right) 
-
\frac{1}{3}\nu\frac{\partial^2 E}{\partial R^2}
.
\label{eeq}
\end{eqnarray}
It is worth noting that (\ref{eeq}) agrees exactly with
Eq.~(58) of \citet{2001MNRAS.325..231O} in the case of a
pressureless viscous fluid with constant kinematic viscosity $\nu$.

To summarise, according to (\ref{tomvan}), all quantities up to second
order are independent of the dynamic time~$t$. The 2nd order
quantities depend on the azimuthal angle $\varphi$ in a linear
combination of 
harmonic oscillations with frequencies 
$m\varphi$, $\tilde m\varphi$ and $2m\varphi$, where
$m = \pm 1$ and $\tilde m = \pm
1$. Therefore, if a non-axisymmetric perturbation occurs, it has to be
a superposition of one-armed and two-armed spiral
structures in second order. 
This is an effect of the non-linearity of the analysis.

\section{Local stability}
\label{sec:local}

The solution of (\ref{eeq}) rules the stability or instability of the
viscously spreading ring. Assuming that short wavelengths become
first unstable (which is supported by the numerical results), we make
the approach
\begin{equation}
E(\tau,R) = \Re[E_0\,\mathrm{e}^{\textstyle i(kR - \sigma\tau)}]
\quad,
\end{equation}
where $E_0$ is a constant. Then, (\ref{eeq}) gives the dispersion
relation
\begin{equation}\label{disprel}
\sigma = - k V_2 + i \left(\frac{V_1}{R} +
k^2\frac{\nu}{3}\right) 
\end{equation}
with
\begin{equation}\label{pe1}
V_1(\tau,R) = 
\frac{1}{2}\frac{\partial\nu}{\partial R} + 
\frac{3}{2}R\frac{\partial^2\nu}{\partial R^2} -
\frac{\nu}{\hat\Sigma_0}\frac{\partial\hat\Sigma_0}{\partial R} +
\frac{3}{2}\frac{R}{\hat\Sigma_0}
\frac{\partial\nu}{\partial R}
\frac{\partial\hat\Sigma_0}{\partial R}
\end{equation}
and
\begin{equation}\label{pe2}
V_2(\tau,R) = 
\frac{\nu}{R} + 
\frac{25}{6}\frac{\partial\nu}{\partial R} +
\frac{8}{3}\frac{\nu}{\hat\Sigma_0}
\frac{\partial\hat\Sigma_0}{\partial R}
\end{equation}
which are real functions. As soon as the imaginary part of $\sigma$
becomes positive, 
\begin{equation}
\sigma_i = \frac{V_1}{R} + k^2\frac{\nu}{3} > 0
\quad,
\end{equation}
the ring becomes unstable. 
This inequality implies that the axisymmetric ring is always
unstable to perturbations of sufficiently short wavelength.
Since there exists also a non-vanishing
real part of $\sigma$, which evokes oscillatory behaviour, and because
the instability is driven by the kinematic viscosity $\nu$, it
is a type of viscous overstability \citep[see e.g.][]{1993ApJ...409..739K}.

Assume $\nu = \mbox{const.}$ Then, the dispersion relation
(\ref{disprel}) takes the form
\begin{equation}
\sigma = 
-k\left(\frac{\nu}{R} + \frac{8}{3}\frac{\nu}{\hat\Sigma_0}
\frac{\partial\hat\Sigma_0}{\partial R}
\right) 
+
i\left(k^2\frac{\nu}{3} - 
\frac{\nu}{R\hat\Sigma_0}\frac{\partial\hat\Sigma_0}{\partial R}
\right)
\;,
\end{equation}
and the condition for instability is
\begin{equation}
k^2 > \frac{3}{R\hat\Sigma_0}\frac{\partial\hat\Sigma_0}{\partial R}
\quad.
\end{equation}

Using the analytic solution (\ref{asd}) of the surface density
$\hat\Sigma_0$, the growth rate $\sigma_i$ becomes
\begin{equation}
\sigma_i =
\frac{1}{3}k^2\nu
+
\frac{\nu}{2R_0^2x^2\tau^*}
\left[
(\tau^* + 4x^2)
-4x \frac{I_{-\frac{3}{4}}\!\!\left(\frac{2x}{\tau^*}\right)}
{I_{\frac{1}{4}}\!\!\left(\frac{2x}{\tau^*}\right)}
\right]
\quad.
\end{equation}
Therefore, with the assumption of small viscous times, i.e.\
$\tau^*\ll x$, the growth rate is
\begin{equation}\label{grst}
\sigma_i = 
\frac{1}{3}k^2\nu +
\frac{3}{4}\frac{\nu}{R^2} + \frac{1}{6\tau}\left(
1 - \frac{R_0}{R}
\right)
\quad,
\end{equation}
where suitable approximations of the modified Bessel functions have
been applied. The condition for instability is accordingly
\begin{equation}
k^2 
 >
- \frac{9}{4 R^2} - \frac{1}{2\nu\tau}\left(
1 - \frac{R_0}{R}
\right)
\quad.
\end{equation}

With the assumption of large viscous times, i.e.\ $\tau^* \gg x$, the
growth rate becomes
\begin{equation}
\sigma_i = \frac{1}{3}k^2\nu + \frac{1}{6\tau}
\quad.
\end{equation}
Obviously, for late times the viscous ring is unconditionally unstable
against non-axisymmetric perturbations.

Although according to the dispersion relation the instability should
grow faster with increasing wave number, in the numerical simulations
finite wavelengths dominate the perturbation. There are two reasons
why the instability might not occur at very short wavelengths. First,
the relevant scales may not be resolved in the numerical calculations.
This effect is indicated in Fig.~\ref{fig:ekin_r}, which shows that
the instability grows faster with higher spatial resolution of the
numerical method. Secondly, the analysis will break down when the
separation of dynamical and viscous timescale becomes invalid, i.e.,
below the characteristic radial scale $R\sim(\nu/\Omega)^{1/2}$.

The main limitation of the presented analysis is the omission of
pressure. In particular, the more general analysis of
\citet{2001MNRAS.325..231O} shows that, in a disc with significant
pressure, the eccentricity vector precesses rapidly and differentially
due to pressure, in addition to its viscous growth or decay. When
pressure dominates over viscosity, the afore mentioned characteristic
radial scale where the analysis may break down is replaced by the
semi-thickness of the disc. Furthermore, the eccentric instability
can be enhanced or suppressed by allowing for the kinematic viscosity
to depend on the surface density, by introducing a bulk viscosity, by
allowing for stress relaxation, or by including three-dimensional
effects.  

\section{Numerical models}
\label{sec:numerics}
To verify the theoretical results obtained in
Sects.~\ref{sec:perturbation} 
and \ref{sec:local}, we performed various 
numerical simulations of the viscously evolving ring, concentrating on
the case of constant kinematic viscosity $\nu_s = \mathrm{const.}$ To
be able to distinguish between numerical and physical effects, we
selected two fundamentally different numerical methods for the
simulations, and we inspected a wide range of parameters.
\subsection{Using SPH}
The first method we used was smoothed particle
hydrodynamics (SPH). This Lagrangian particle method was introduced by
\citet{1997MNRAS...181..375G} and \citet{1977AJ...12..1013L} to
simulate compressible flows with free boundaries. For an overview of
the SPH method see, for example, \citet{1992ARAA...30..543M}.

In our simulations we did not use the artificial viscosity approach
\citep[see e.g.][]{1983JCoP..52..374M} usually applied to model viscid
fluids with SPH, but we chose a version of SPH that includes the
entire viscous stress tensor
\citep{1994ApJ...431..754F,R1995CPC..89..1R}. This implementation
allows for the proper treatment of viscous shear flows. Additionally,
in compliance with the assumption of a cold disc, pressure was
completely neglected in all the simulations. A detailed description of
the application of this SPH approach to the viscously evolving ring
can be found in \citet{Spei99}.

%%%%
\begin{figure*}
\begin{minipage}{0.32\textwidth}
\resizebox{\hsize}{!}
{\includegraphics{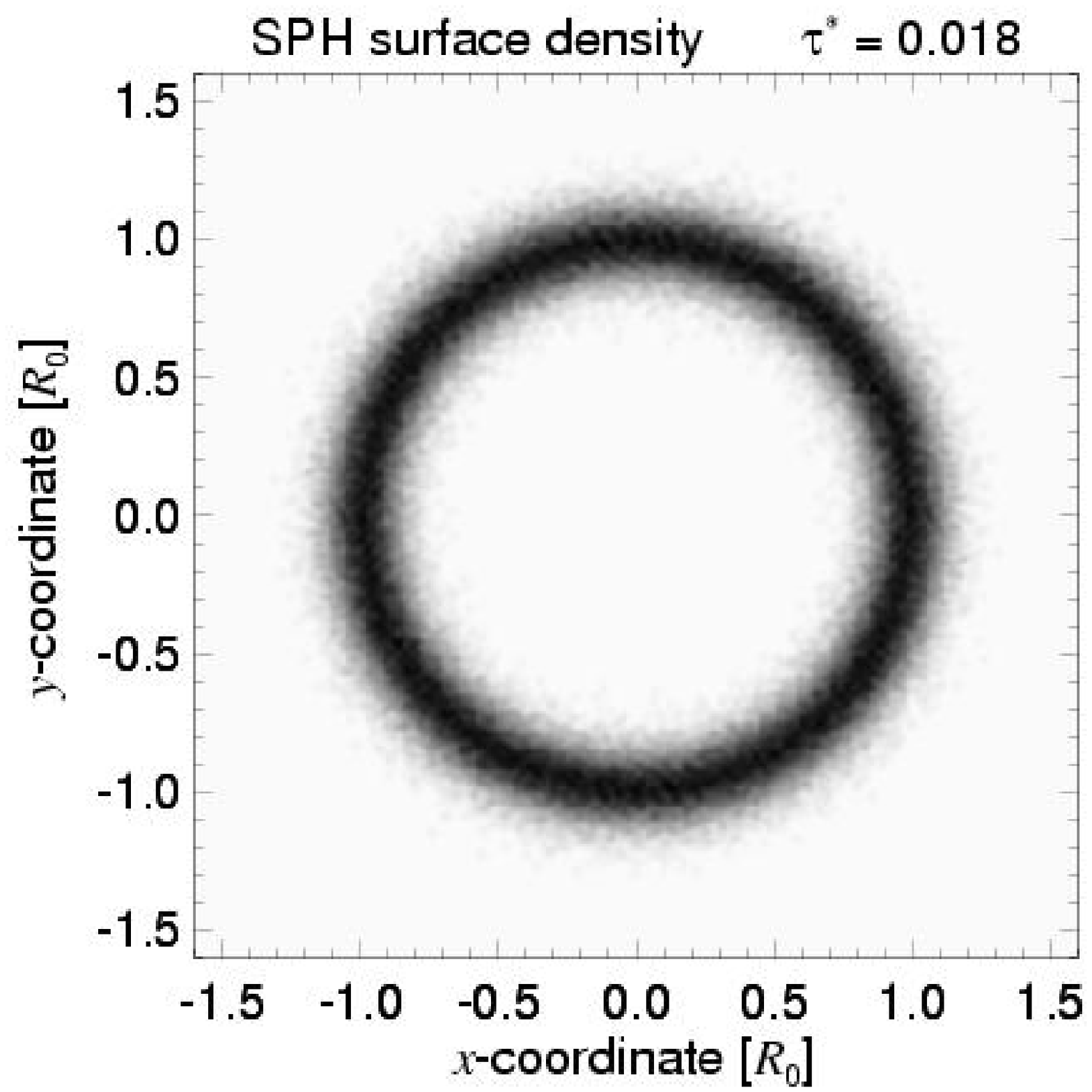}}
\end{minipage}
\hfill
\begin{minipage}{0.32\textwidth}
\resizebox{\hsize}{!}
{\includegraphics{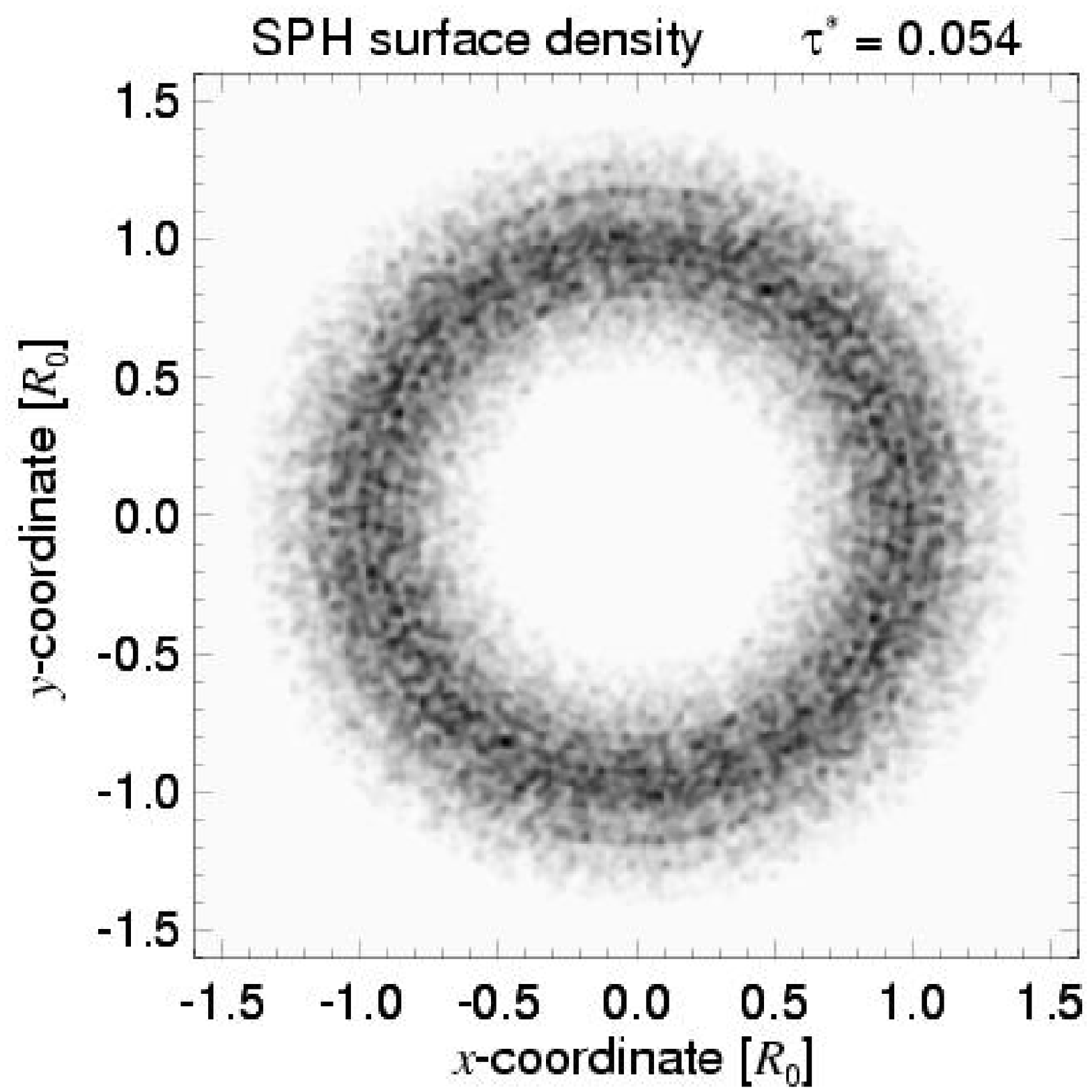}}
\end{minipage}
\hfill
\begin{minipage}{0.32\textwidth}
\resizebox{\hsize}{!}
{\includegraphics{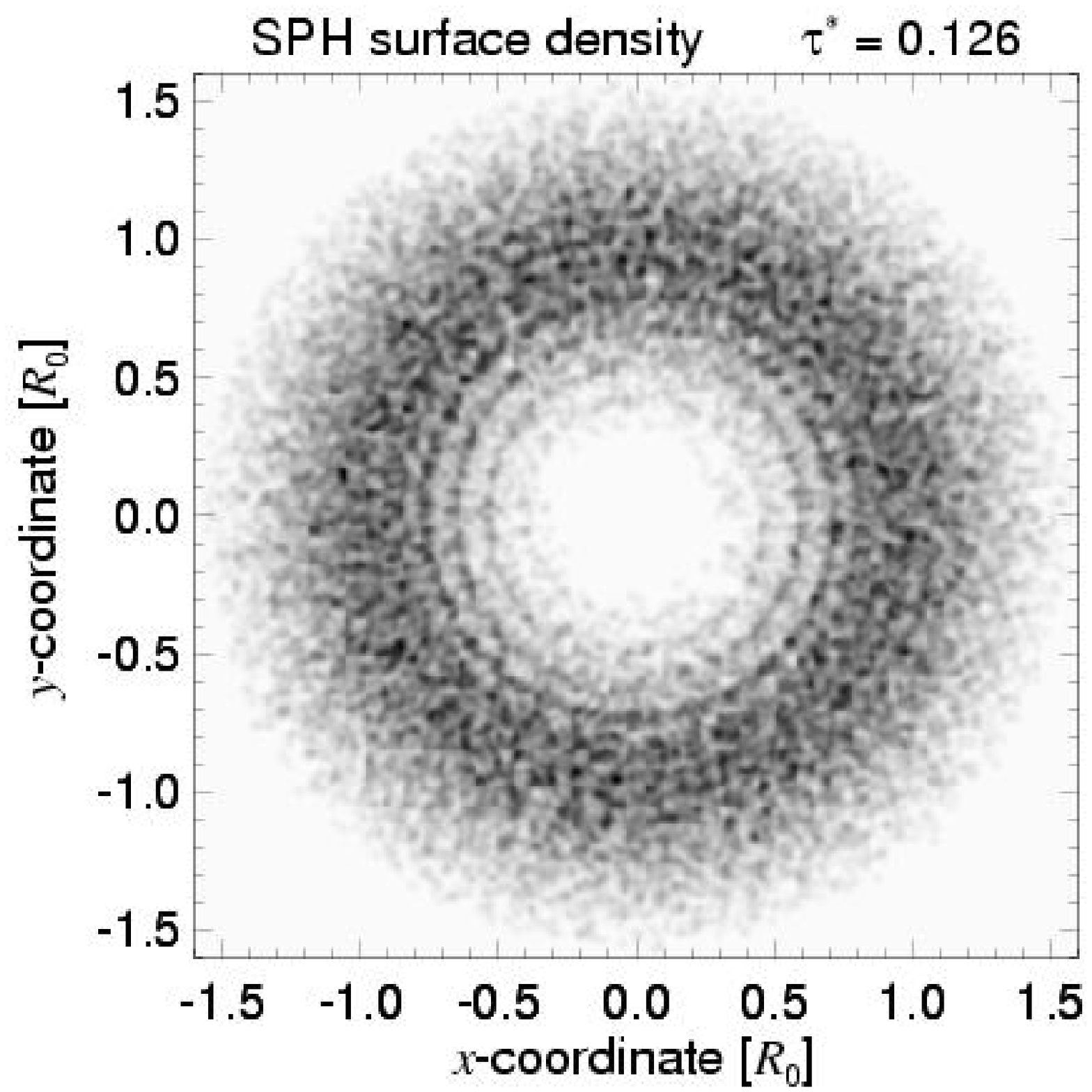}}
\end{minipage}
\\[2ex]
\begin{minipage}{0.32\textwidth}
\resizebox{\hsize}{!}
{\includegraphics{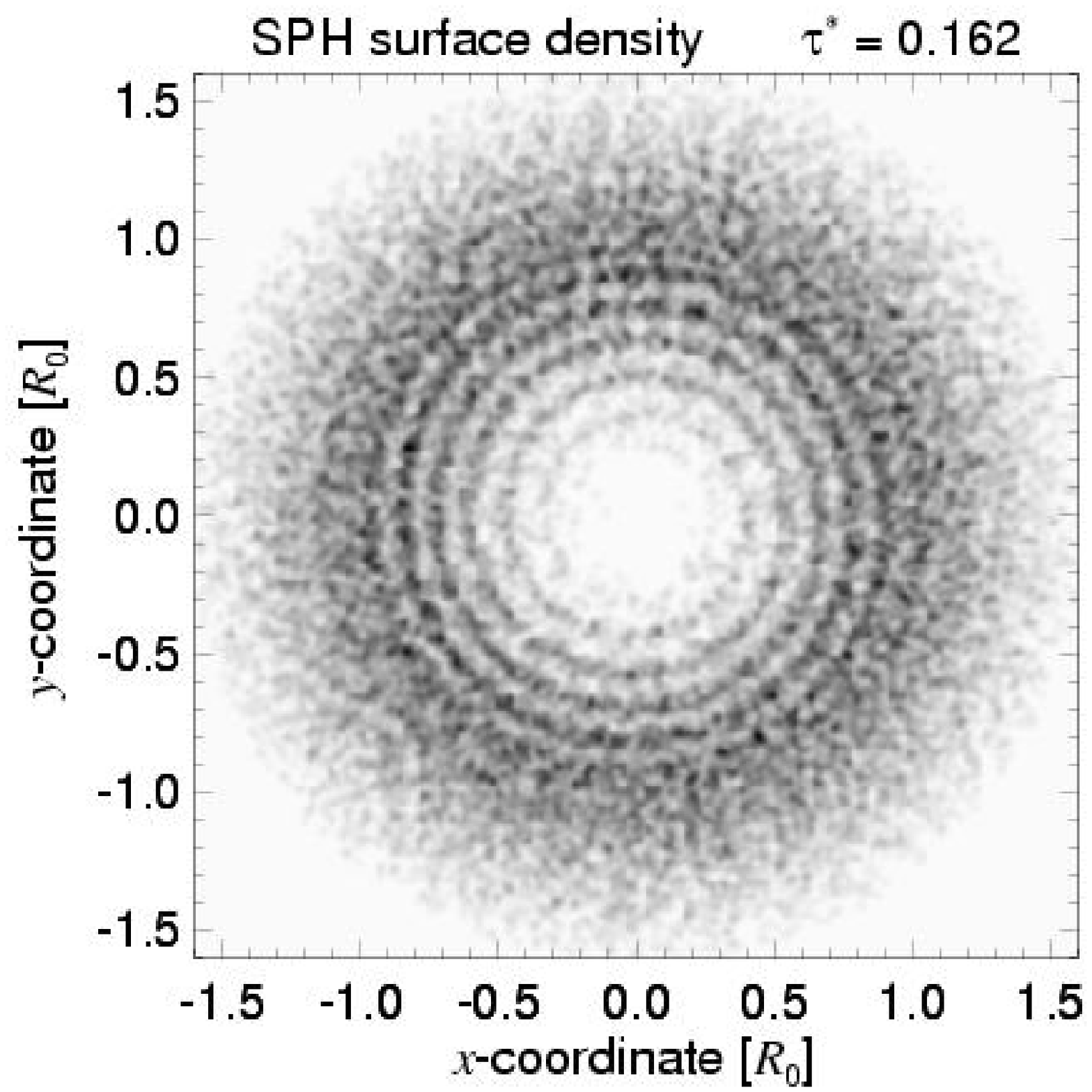}}
\end{minipage}
\hfill
\begin{minipage}{0.32\textwidth}
\resizebox{\hsize}{!}
{\includegraphics{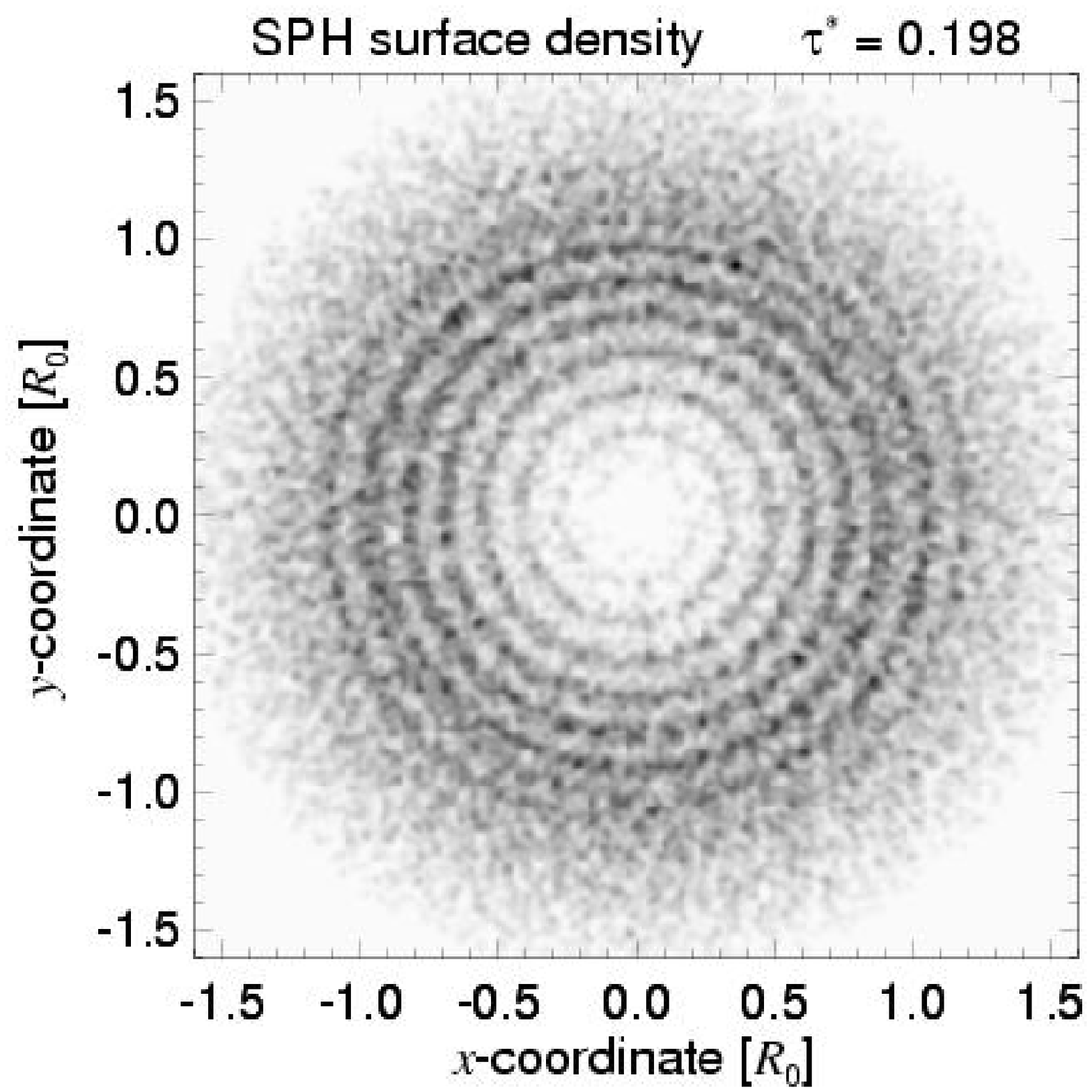}}
\end{minipage}
\hfill
\begin{minipage}{0.32\textwidth}
\resizebox{\hsize}{!}
{\includegraphics{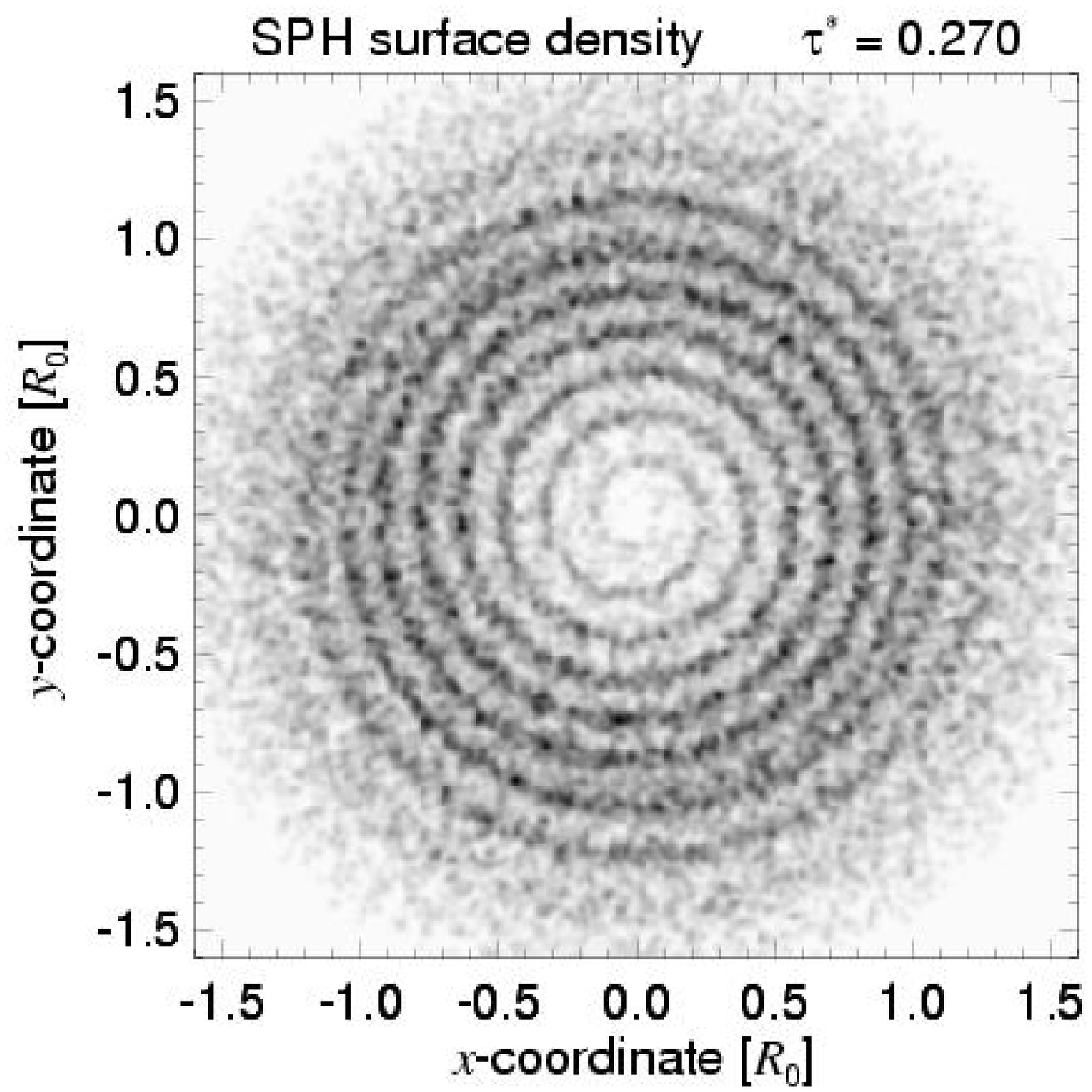}}
\end{minipage}
\caption{\label{fig:sph-densevol} 
Evolution of the surface density of the viscous dust ring in an
SPH simulation. Parameters of the disc are $R_0 = 1\,R_{\odot}$, disc
mass $M = 10^{-10}\,M_\odot$, central mass $M_\mathrm{c} =
1\,M_\odot$, and kinematic viscosity $\nu_s = 3\times 10^{-8}
R_{\odot}^2/\mathrm{s} = 1.5\times 10^{14} \mathrm{cm}^2/\mathrm{s}$.
The initial density distribution at viscous time $\tau^* = 0.018$ is
shown in the top left plot. The symmetric structures seen in the
distribution at $\tau^* = 0.054$ may be of numerical nature due to
relaxation effects of the initial particle distribution. 
Later, the development of the spiral instability can be
seen very clearly.
}
\end{figure*}
%%%%

The simulation presented below was performed with the following
parameters. The initial SPH particle distribution represents the
surface density (\ref{asd}) of the analytic solution at the viscous
time $\tau^* = 0.018$, with $R_0 = 1\,R_{\odot}$, disc mass $M =
10^{-10}\,M_\odot$, and central mass $M_\mathrm{c} = 1\,M_\odot$. The
kinematic viscosity was set to $\nu_s = 3\times 10^{-8}
R_{\odot}^2/\mathrm{s} = 1.5\times 10^{14} \mathrm{cm}^2/\mathrm{s}$
throughout the whole simulation. The initial distribution consists of
40\,000 particles which are arranged symmetrically to make sure that
at the beginning the centre of mass lay exactly at the origin. 
Due to the stochastic nature of the SPH method, no further
seed perturbations are required to bring potential instabilities to
grow.

Fig.~\ref{fig:sph-densevol} shows the evolution of the viscous ring
during the SPH simulation in a sequence of plots of the surface
density distribution. The top left plot represents the initial density
distribution at viscous time $\tau^* = 0.018$. Immediately after the
start of the simulation, symmetric structures develop, which can still
be seen in the distribution at $\tau^* = 0.054$ in the top middle
panel. These narrow ring-like structures may be numerical
artifacts due to the relaxation of the initial particle distribution,
or they may be caused by the radial viscous overstability discussed
e.g.\ in \citet{1993ApJ...409..739K},
but they vanish in the further course of the simulation. Instead, the
development of a spiral pattern begins, in this case at the inner edge
of the disc, until finally a distinct spiral structure is covering the
whole face of the disc, as can clearly be seen in the distribution at
$\tau^* = 0.270$ in the bottom right panel of
Fig.~\ref{fig:sph-densevol}.

%%%%
\begin{figure*}
\begin{minipage}{0.49\textwidth}
\resizebox{\hsize}{!}
{\includegraphics{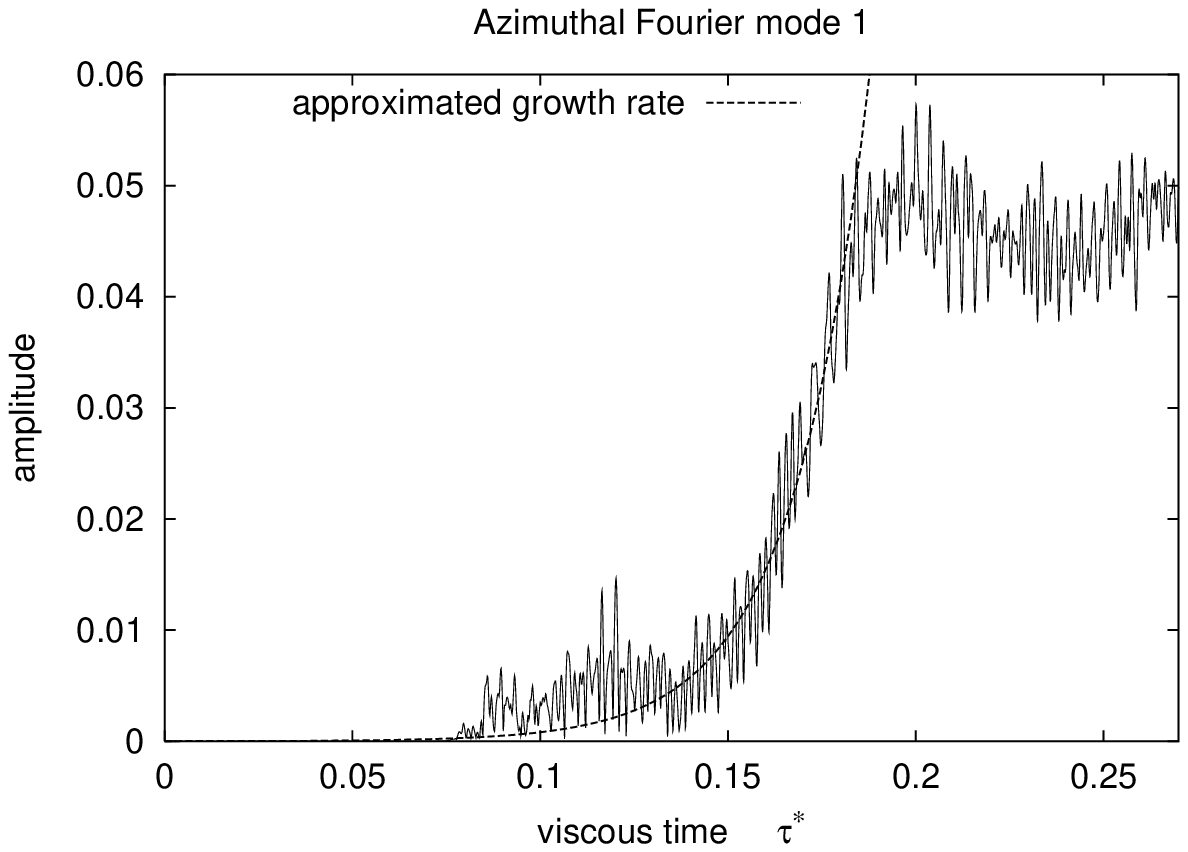}}
\end{minipage}
\hfill
\begin{minipage}{0.49\textwidth}
\resizebox{\hsize}{!}
{\includegraphics{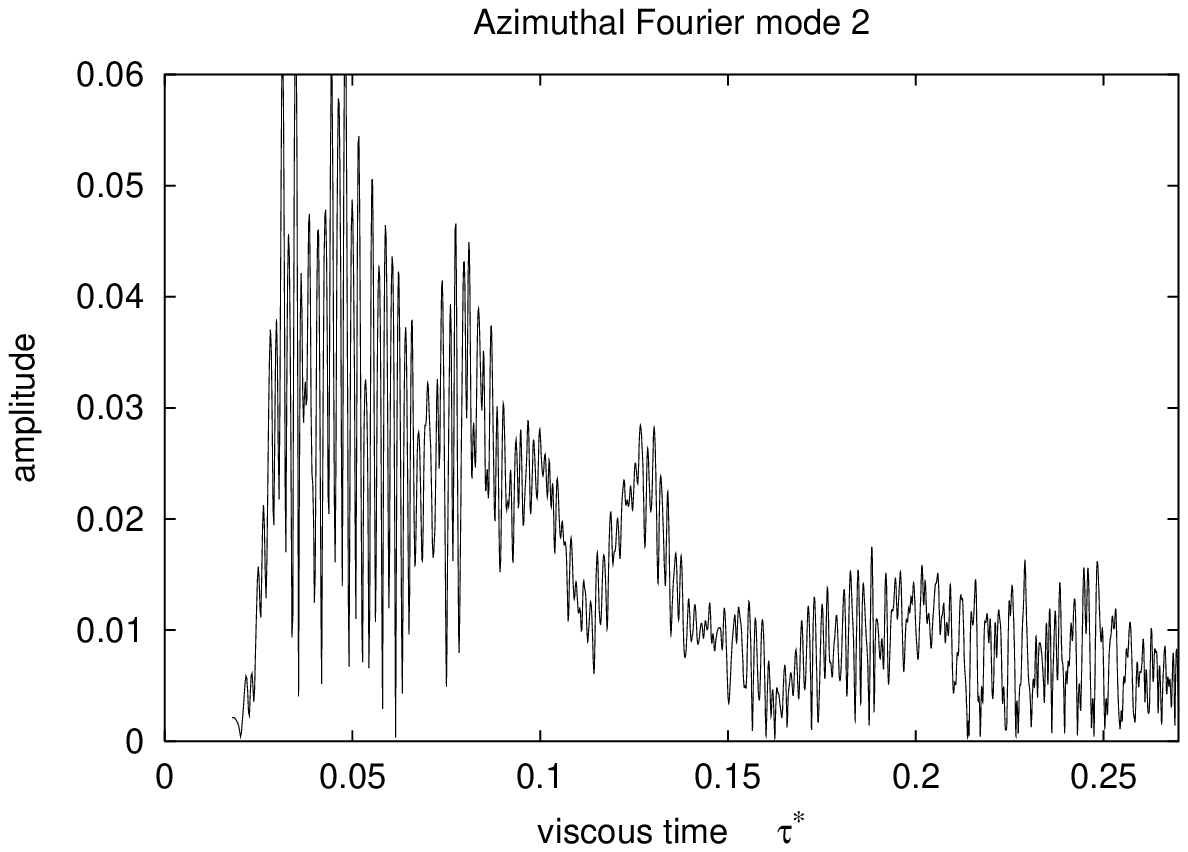}}
\end{minipage}
\\[2ex]
\begin{minipage}{0.49\textwidth}
\resizebox{\hsize}{!}
{\includegraphics{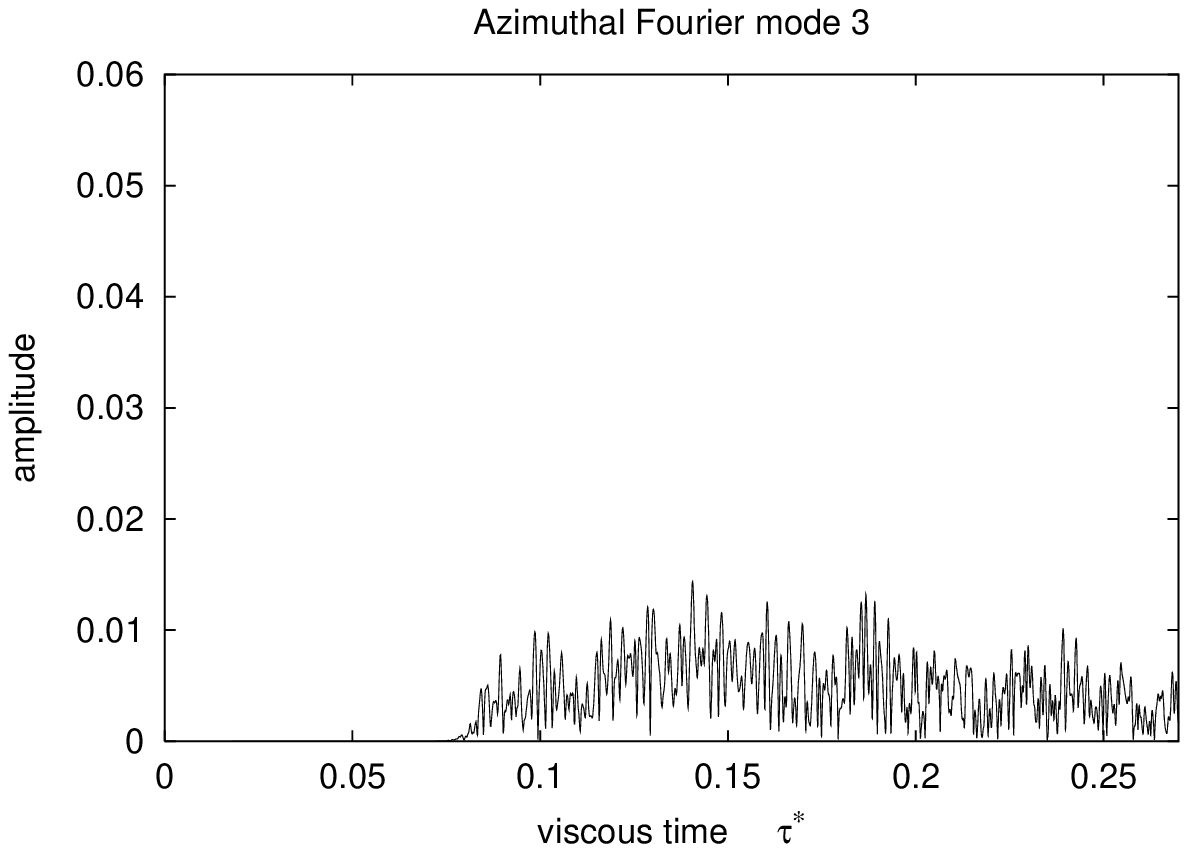}}
\end{minipage}
\hfill
\begin{minipage}{0.49\textwidth}
\resizebox{\hsize}{!}
{\includegraphics{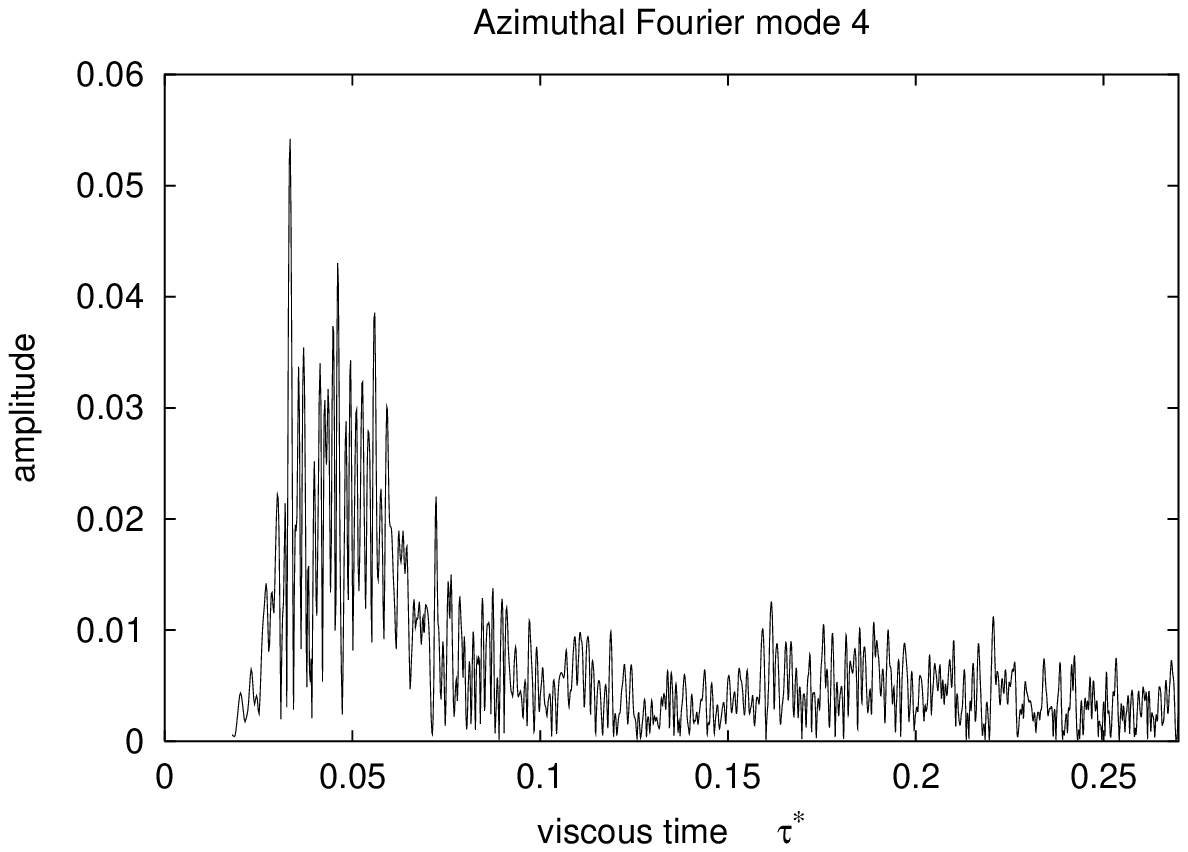}}
\end{minipage}
\caption{\label{fig:sph-modes} 
Time evolution of the first four Fourier modes at $R = R_0$ of the
simulation shown in Fig.~\ref{fig:sph-densevol}. Because of the
symmetric initial particle distribution, even modes are high and odd
modes are low at the beginning. Later, all modes decrease except the
$(m=1)$-mode which increases exponentially as expected. The $(m=3)$-
and $(m=4)$-mode merge with the noise background of the simulation,
while the $(m=2)$-mode stays slightly above the noise level. The
dashed line in the top left panel gives an approximated fit of the
growth of the first Fourier mode to $\sigma_i = 1.8\times
10^{-5} \mathrm{sec}^{-1} = 0.028\,\Omega(R_0)$.
}
\end{figure*}
%%%%

To determine the nature of the spiral structure, a Fourier analysis of
the evolving ring was performed during the simulation.
Fig.~\ref{fig:sph-modes} gives the time evolution of the first four
azimuthal Fourier modes ($m = 1$ to $m = 4$) at radius $R = R_0$. For
small viscous times, the even modes $m = 4$ and especially $m = 2$ are
high, while the odd modes $m = 1$ and $m = 3$ are low. This effect is
due to the symmetric initial particle distribution and disappears
eventually when the symmetry of the particle distribution is lost.
Later, the ($m=1$)-mode increases exponentially, while all the other
modes decrease or stay small. In particular, the ($m=3$)-mode remains
in the large noise background of the simulation, which is intrinsic to
the SPH method, and the ($m=4$)-modes decreases down to the noise
background (as all inspected higher modes do). The $(m=2)$-mode seems
to end at a level slightly above the noise background, although this
is not clearly to determine because of the large oscillations of the
modes which are also caused by the SPH noise level.

The results of the Fourier analysis, i.e., the development of a
dominant one-armed spiral structure with an additional weaker component
of a two-armed spiral structure, are in complete agreement with the
theoretical results of the first and second order perturbation
analysis as derived in \ref{sec:pert1o} and \ref{sec:pert2o}.

%%%%
\begin{figure}
\resizebox{\hsize}{!}
{\includegraphics{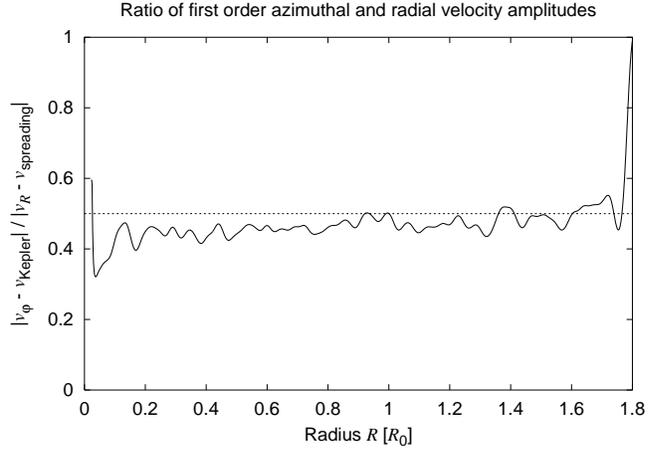}}
\caption{\label{fig:sph-velratio} 
Ratio of the amplitudes of the errors of azimuthal and radial
velocities at $\tau^* = 0.270$ for the disc shown in the bottom right
panel of Fig.~\ref{fig:sph-densevol}. Plotted is the azimuthally
averaged relation $|v_{\varphi} - v_\mathrm{Kepler}|/|v_{R} -
v_\mathrm{spreading}|$ over radius, where $v_{\varphi}$ and $v_{R}$
are the simulation results and $v_\mathrm{Kepler}$ and
$v_\mathrm{spreading}$ are the velocities (\ref{vkepler}) and
(\ref{usol1}) of the analytic solution. The curve roughly matches the
expected value of $\frac{1}{2}$ (dotted line).
}
\end{figure}
%%%%

%%%%
\begin{figure}
\resizebox{\hsize}{!}
{\includegraphics{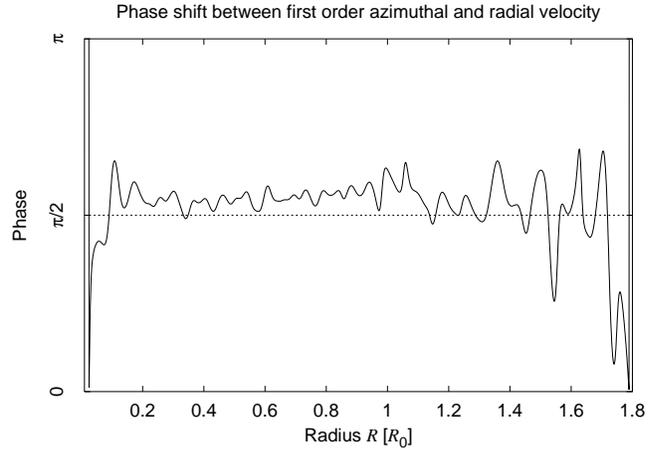}}
\caption{\label{fig:sph-phshift}
Absolute value of the phase shift between the errors of azimuthal and
radial velocities, $v_{\varphi} - v_\mathrm{Kepler}$ and $v_{R} -
v_\mathrm{spreading}$, for the disc at $\tau^* = 0.270$ shown in the
bottom right panel of Fig.~\ref{fig:sph-densevol}, azimuthally
averaged and plotted over radius. The curve roughly matches the
expected value of $\frac{\pi}{2}$ (dotted line).
}
\end{figure}
%%%%

Further predictions of the perturbation analysis in 
Sect.~\ref{sec:perturbation}, that can be tested easily, are the relations
between the first order velocities $v_{R1}$ and $v_{\varphi
1}$. According to (\ref{factor2}), their amplitudes have a ratio of 2,
and they obey a phase shift of $\frac{\pi}{2}$.  To verify this, we
more thoroughly studied the most evolved particle distribution of the
SPH simulation presented in Fig.~\ref{fig:sph-densevol}, i.e.\ the
ring at viscous time $\tau^* = 0.270$, whose density distribution is
shown in the bottom right panel of Fig.~\ref{fig:sph-densevol}. In
Fig.~\ref{fig:sph-velratio} the relation $|v_{\varphi} -
v_\mathrm{Kepler}|/|v_{R} - v_\mathrm{spreading}|$ is plotted over
radius. Here, $v_{\varphi}$ and $v_{R}$ denote the simulation results,
which are averaged azimuthally using the SPH formalism, and
$v_\mathrm{Kepler}$ and $v_\mathrm{spreading}$ are the velocities
according to the analytic solutions (\ref{vkepler}) and (\ref{usol1})
of the viscous ring model. Taking $v_{\varphi} - v_\mathrm{Kepler}
\approx v_{\varphi 1}$ and $v_{R} - v_\mathrm{spreading} \approx
v_{R1}$, the curve in diagram~\ref{fig:sph-velratio} matches the value
$\frac{1}{2}$ sufficiently to satisfy relation
(\ref{factor2}). Accordingly, Fig.~\ref{fig:sph-phshift} shows the
phase shift between $v_{\varphi} - v_\mathrm{Kepler}$ and $v_{R} -
v_\mathrm{spreading}$, and again the plot matches the expected value
of $\frac{\pi}{2}$ satisfactorily.

%%%%
\begin{figure}
\resizebox{\hsize}{!}
{\includegraphics{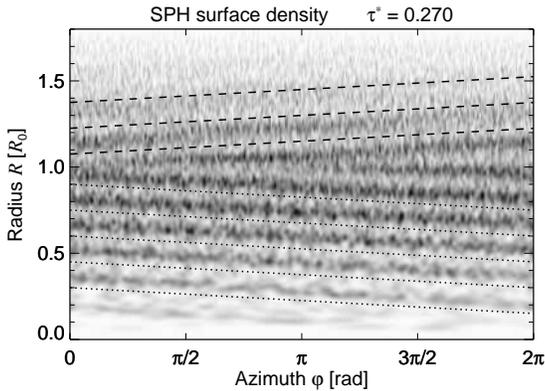}}
\caption{\label{fig:sph-phirdens} 
Surface density distribution of the disc at $\tau^* = 0.270$ as shown
in the bottom right panel of Fig.~\ref{fig:sph-densevol} but plotted
in polar coordinates (radius $R$ over azimuth $\varphi$) instead of
Cartesian coordinates. Additionally, lines of constant $kR \pm
\varphi$ with $k = 42 R_0^{-1}$ are drawn in around $R \approx
R_0$. Note that the spiral structure consists of a leading and a
trailing spiral.
}
\end{figure}
%%%%

Finally, we want to compare the local stability analysis and the
resulting dispersion relation established in Sect.~\ref{sec:local}
with the SPH simulation results. From the top left panel of
Fig.~\ref{fig:sph-modes}, the growth rate at radius $R = R_0$ of 
the first Fourier mode may be estimated to $\sigma_i = 1.8\times
10^{-5} \mathrm{sec}^{-1} = 0.028\,\Omega(R_0)$, the latter measured
in units of the angular velocity (\ref{angvel}) at $R = R_0$. With dispersion
relation (\ref{grst}), the corresponding wave number can be estimated
to $k = 42 R_0^{-1}$, which results in a wavelength of $\lambda =
0.15 R_0$. Comparing that result with the surface density
distributions of Fig.~\ref{fig:sph-densevol} shows good agreement.
That can be seen in particular in Fig.~\ref{fig:sph-phirdens}, where
the surface density distribution of the ring at viscous time $\tau^* =
0.270$ is plotted in polar coordinates, and where additionally lines
of constant $kR \pm \varphi$ with $k = 42 R_0^{-1}$ are drawn in for
the region around $R \approx R_0$. The lines match well the slopes and
the wavelengths of the spiral structures. Note by the way that in this
presentation of the surface density clearly can be seen that in this
case the spiral structure consists of a leading (outer part of the
disc) and a trailing (inner part of the disc) spiral.

\subsection{Using a finite difference method}
\subsubsection{Numerical considerations}
In addition to the particle simulations above we performed
runs using a finite difference method for solving the hydrodynamic
equations.
The code is based on the hydrodynamic program {\it RH2D}
suited to study general two-dimensional systems including
radiative transport \citep{1989A&A...208...98K}.
{\it RH2D} uses a fixed Eulerian grid and is
a spatially second order accurate, mixed explicit
and implicit method.
Due to an operator-splitting approach, the method is partly centred in
time and is therefore also in time in real terms of higher accuracy than
the formal first order.
The advection is computed by means of the second order monotonic
transport algorithm, introduced by \citet{1977JCoPh..23..276V},
which guarantees global conservation of mass and angular momentum.
Advection and forces are solved explicitly, while the
viscosity is treated either explicitly or implicitly.
The formalism for application to thin discs in $(R -\varphi)$ geometry
has been described in detail in
\citet{1998A&A...338L..37K,1999MNRAS.303..696K}.
 
To model the viscously evolving ring we work in cylindrical coordinates
and solve exactly the hydrodynamic equations as given in Eqs.~(\ref{ceq})
to (\ref{weq}), using the full viscous stress tensor presented
in (\ref{shrr}) to (\ref{shpp}), using an extremely small pressure with
$c_\mathrm{s}/v_\mathrm{Kepler} \approx 10^{-8}$.
The equations are solved in dimensionless units using an arbitrary
basic unit length $R_0$.
The unit of time $t_0$ is chosen such that
\begin{equation}
          t_0 = \sqrt{\frac{R_0^3} {G M_\mathrm{c}} }
\quad,
\end{equation}
i.e.\ in these units $G M_\mathrm{c} =1$, and the period of one
Keplerian orbit at (the dimensionless radius) $R=1$ is $P_0 = 2 \pi$.
The only relevant physical constant entering the equations is the
dimensionless coefficient of the kinematic viscosity $\nu_{s}$,
which is measured in units of $\nu_0 = R_0^2 / t_0$.

All the grid-based models presented in this paper use the same computational
domain ${\cal D}$ represented by $[R_\mathrm{min}, R_\mathrm{max}] \times 
[\varphi_\mathrm{min}, \varphi_\mathrm{max}]$ with
$R_\mathrm{min} = 0.2$, $R_\mathrm{max} = 1.8$, $\varphi_\mathrm{min} = 0.0$
and $\varphi_\mathrm{max} = 2 \pi$.  
The ring ${\cal D}$ is covered in the basic model by $128 \times 128$
grid cells. However, to analyse resolution and numerical 
effects we use different griddings ranging from
$64 \times 128$ to $256 \times 733$.  
For our basic reference model we use for the constant dimensionless
viscosity $\nu_{s} = 4.77\times 10^{-5}$, which is in fact identical
to the value used in the particle simulations.
\begin{figure*}
\begin{center}
\resizebox{0.9\linewidth}{!}{%
\includegraphics{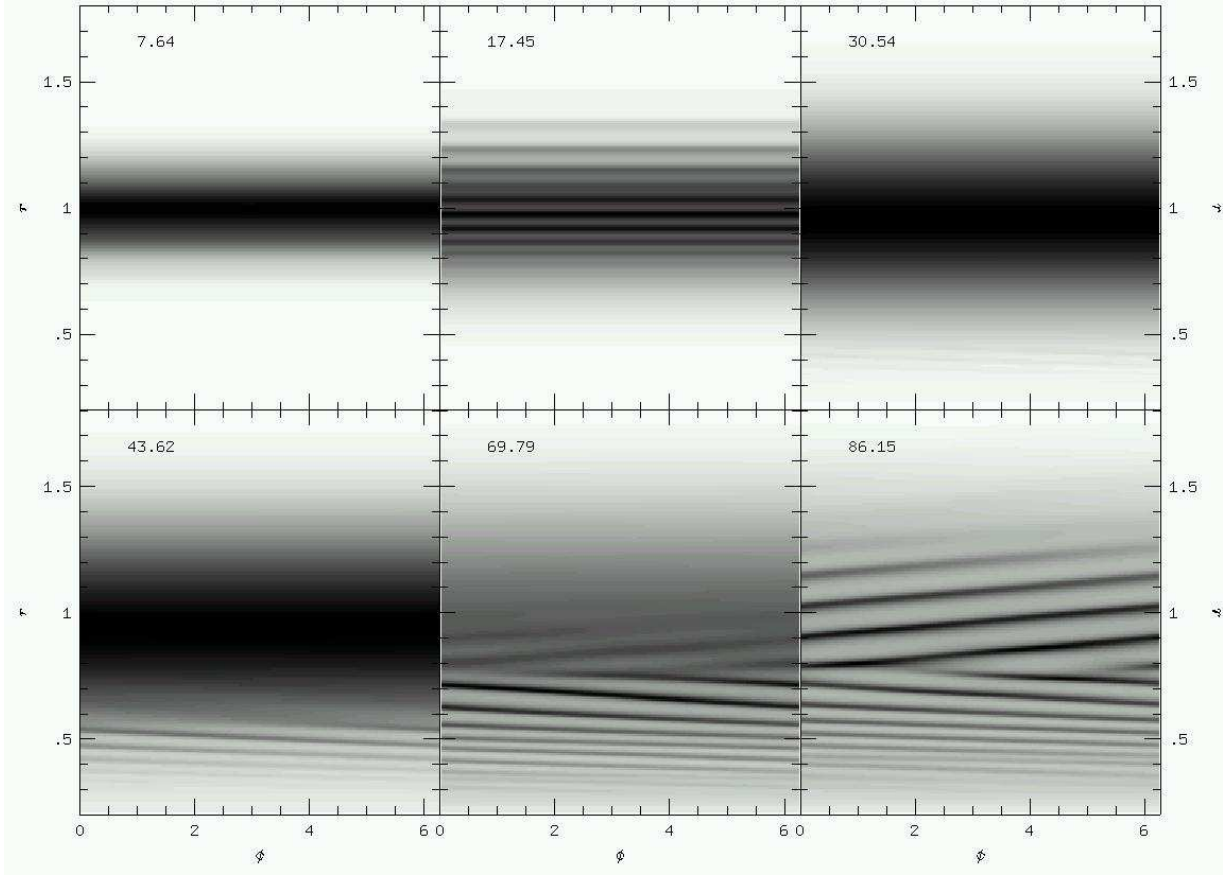}}
\end{center}
  \caption{Grey-scale plot of the density in polar coordinates
  at different evolutionary times
  for a model with a $256 \times 256$ resolution. Times are indicated
  in orbital periods $P_0$ at $R=1 R_0$. To compare with 
  Fig.~\ref{fig:sph-densevol}: One viscous time refers to $278 P_0$.
  The time evolution of the system proceeds similar to the particle
  simulations. In the initial phase axisymmetric perturbations
  are seen first ($t=17.45$). They vanish later, and the system returns to
  axisymmetry ($t=30.54$). Then, from the inner parts the
  instability develops as a {\it trailing} spiral. At the end of the
  simulation a {\it leading} spiral appears in the outer region of the ring.
   \label{fig:ring5s.rp6} 
    }
\end{figure*}

The initial setup consists of an axisymmetric density
profile following the analytic density
$\Sigma_\mathrm{ring}$ as given in Eq.~(\ref{asd}).
Because the original $\delta$-function distribution is hard
to represent numerically, we start, as above,
with an initial density profile given here by the
analytic solution at the viscous time $\tau^* = 0.016$, i.e.
$\Sigma (t = t_\mathrm{init}, x = R/R_0) = 
\Sigma_\mathrm{ring} (\tau^* = 0.016, x)$.
The slight difference to the value 0.018, used in the SPH calculations,
is of no importance for the subsequent evolution.
For the standard viscosity $4.77\times 10^{-5}$ this refers to the initial
(dimensionless) time 
$t_\mathrm{init} = \tau^* R_0^2 / (12 \nu_\mathrm{s}) = 4.34 P_0$.
Stated differently, one viscous evolution time corresponds
to 278 orbital periods $P_0$ for the given viscosity.
The total dimensionless mass $M$ in the disc is normalised to unity.
The initial radial velocity is set to zero and the angular viscosity
to $v_\mathrm{Kepler}$.

The code is written such that axial symmetry is preserved exactly 
for an {\it explicit} viscosity-solver, i.e. purely
axisymmetric initial conditions remain exactly axisymmetric. Hence, the
initial density is disturbed randomly by typically $1$ or $0.1\%$ 
to supply seed perturbations,
and the subsequent
longterm integration has to follow the evolution on a viscous timescale
which is typically about $\approx 100$ orbital periods for the standard model.
\subsubsection{The standard model}
An analysis of the properties of the standard model is presented.
This refers to fixed physical parameter and initial conditions 
as outlined in the previous section.
Starting from the disturbed axisymmetric density distribution
the configuration evolves 
as presented
in Fig.~\ref{fig:ring5s.rp6}
where the density distribution is plotted for a model with a resolution
of $256 \times 256$ and an initial density perturbation of $0.1 \%$.
During the initial phase axisymmetric disturbances appear which vary
on short dynamical timescales. 
These radial oscillations may be related to the viscous overstability
discussed for example in 
\citet{1978MNRAS.185..629K}
or in
\citet{1993ApJ...409..739K}.
The oscillations are damped however,
and the system becomes axially symmetric again.
Apart from these initial variations, there are no further 
indications of a variation on the dynamical timescale during the 
remaining evolution.
At $t \approx 35$ a trailing one-armed spiral density wave becomes visible
in the inner parts of the ring (at $R \approx 0.4 R_0$)
which propagates slowly outwards. At later times a leading spiral appears
in the outer parts of the ring which has a negative density gradient.
This is in very good agreement with the SPH simulations presented above,
and with the analytical results which indicated the possibility
of leading and trailing waves.
The outer leading spiral is not as tightly wound as the trailing
inner one. That is, the radial wave number of the outer wave is smaller
than the one on the inner side. This different behaviour may be caused by a
wrapping up of the inner
spiral by the differential rotation.
At a given radius the radial separation of two wave crests (tightness of the
spiral) is approximately constant with time, while for a given snapshot (time)
there appears to be the tendency for the spirals to widen for larger radii.

\begin{figure}
\begin{center}
%\resizebox{0.9\linewidth}{!}{%
\resizebox{0.86\linewidth}{!}{%
\includegraphics{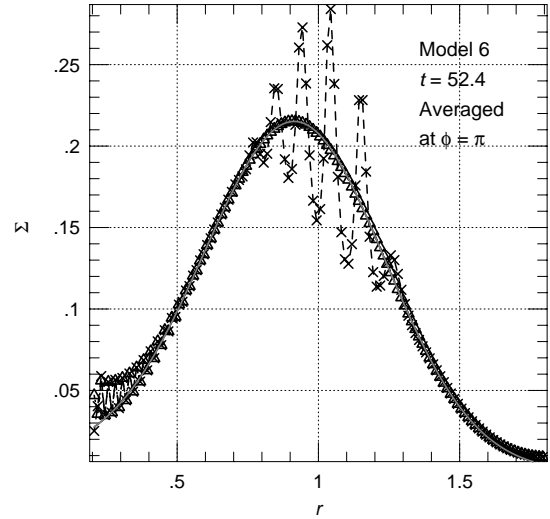}}
\end{center}
  \caption{Radial density distribution at $\varphi = \pi$ (crosses)
   and azimuthally averaged (triangles). The averaged distribution
   follows exactly the analytical zero order result of the
   diffusion equation $\Sigma_\mathrm{ring}$, Eq.~(\ref{asd}).
   Time is given in orbits at $R=1 R_0$.
   \label{fig:ring6.ave}
    }
\end{figure}
In Fig.~\ref{fig:ring6.ave} a radial cut through the density at the
fixed angle $\varphi = \pi$ (crosses) is shown for the standard model
($128 \times 128$ grid points), for $t= 52.4$.
Clearly seen are the different crests of the spirals. The
radial wavelength $\lambda = 2 \pi /k$ of the spiral is of
the order of a few grid points only.
Overlaid we plotted the azimuthally averaged density profile
(triangles) which follows exactly the analytical solution (solid line).
This feature of the averaged density is a result of the modal form
$\propto \, \mathrm{e}^{i m \varphi}$ of the developing spiral structure. 
The oscillations at the inner boundary are caused by the artificial
inner boundary condition which allows for no outflow
($v_R = 0$  at  $R_\mathrm{min}$) for this particular model. Opening
the inner boundary yields smooth density distributions.
The obtained growth rates are, however, not influenced by the exact
boundary conditions, as long as the density there remains small with
respect to the central density.

\begin{figure}
\begin{center}
\resizebox{0.9\linewidth}{!}{%
\includegraphics{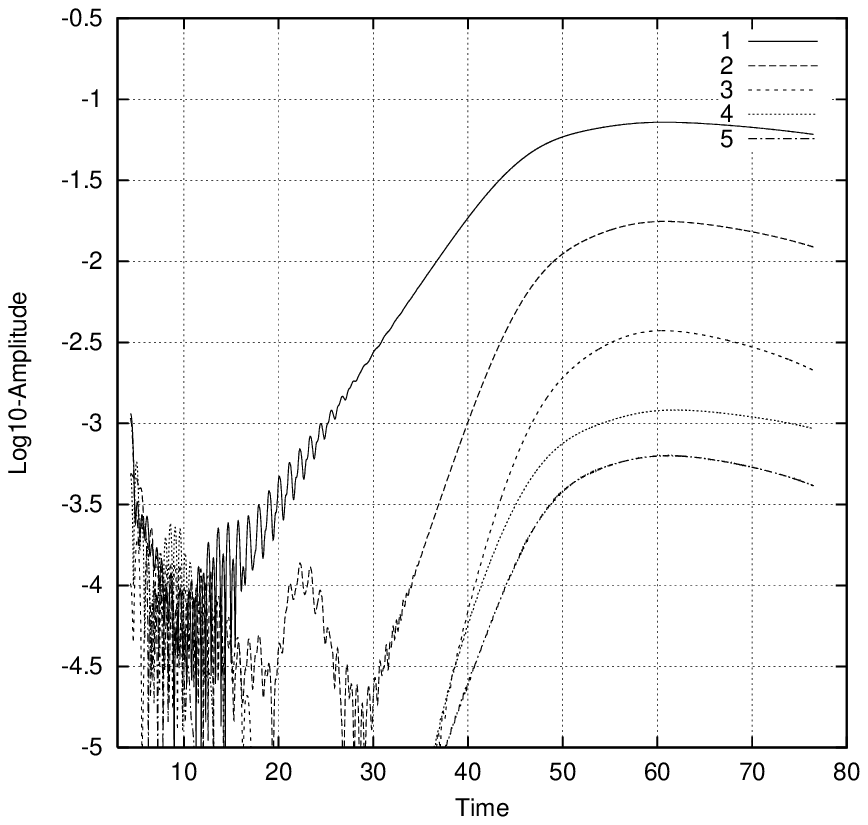}}
\end{center}
  \caption{Time evolution of the first 5 azimuthal modes for the standard
   model measured at the radius $R=1$.
   \label{fig:ring6.modes}
    }
\end{figure}

In Fig.~\ref{fig:ring6.modes} the growth rates for the first 5
modes $m=1$ upto $m=5$ are displayed. The vertical axis refers to
the decimal logarithm of the amplitude. Clearly seen is the random initial
perturbation at the level of $1 \%$ which is reflected in a
start of all modes at the level of $10^{-3}$. During the initial
evolution the rings spreads slowly, the amplitudes decline until
at later times ($t = 15-20$) the $m=1$ mode begins to grow
exponentially with time. From the plot and additional runs
we may approximately determine the growth rate
$\sigma_i = \mathrm{Im}(\sigma)$ for this $m=1$ mode
of the standard model to be 
\begin{displaymath}
                 \sigma_i \approx  0.035 \, \Omega (R=1)\quad.
\end{displaymath}
This is stated in units of the dimensionless time which is identical 
to the Keplerian orbital frequency $\Omega$ at the radius $R=1 R_0 $.
To verify the numerical robustness of this result we varied
several numerical parameter such as rotating frame, implicit viscosity,
directional splitting, inner and outer boundary conditions,
and found no significant variation.

Of course, since the growth rate $\sigma_i = \sigma_i (k)$ depends on the 
azimuthal wave number $k$ and thus on the grid size as well,
we do expect a dependence on the numerical resolution.
\begin{figure}
\begin{center}
\resizebox{0.9\linewidth}{!}{%
\includegraphics{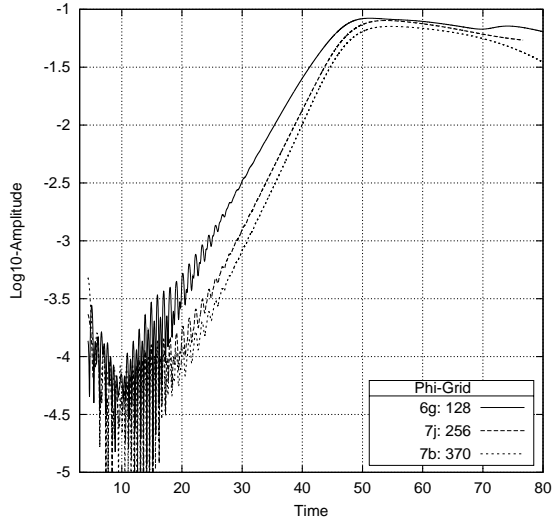}}
\end{center}
  \caption{Time evolution of the logarithm of $m=1$ mode 
   (measured at $R=1 R_0$)
   for models with varying resolution in the angular direction.
   The radial number of grid points is fixed to 128.
   \label{fig:f128}
    }
\end{figure}
\begin{figure}
\begin{center}
\resizebox{0.9\linewidth}{!}{%
\includegraphics{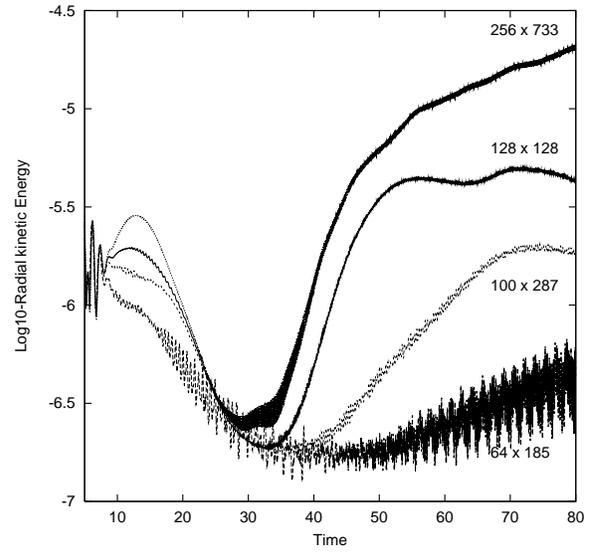}}
\end{center}
  \caption{Time evolution of the total radial kinetic energy
    (dimensionless units)
   in the system for different numerical resolution,
   as indicated by the labels. The physical parameter
   of the models are those of the standard model. 
   \label{fig:ekin_r}
    }
\end{figure}
\begin{figure}
\begin{center}
\resizebox{0.9\linewidth}{!}{%
\includegraphics{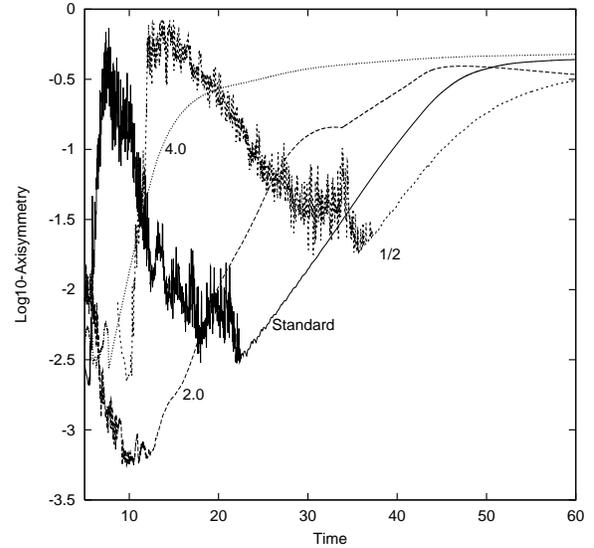}}
\end{center}
  \caption{Time evolution of the asymmetry $\Delta \Sigma$,
   Eq.~(\ref{eq:deltasigma}), for different kinematic viscosities.
   Starting from the reference value (standard, $\nu = 4.77 \times 10^{-5}$)
   we lowered and increased the
   value of $\nu$ by the factors given in the key.
   \label{fig:visc}
    }
\end{figure}
\begin{figure}
\begin{center}
\resizebox{0.9\linewidth}{!}{%
\includegraphics{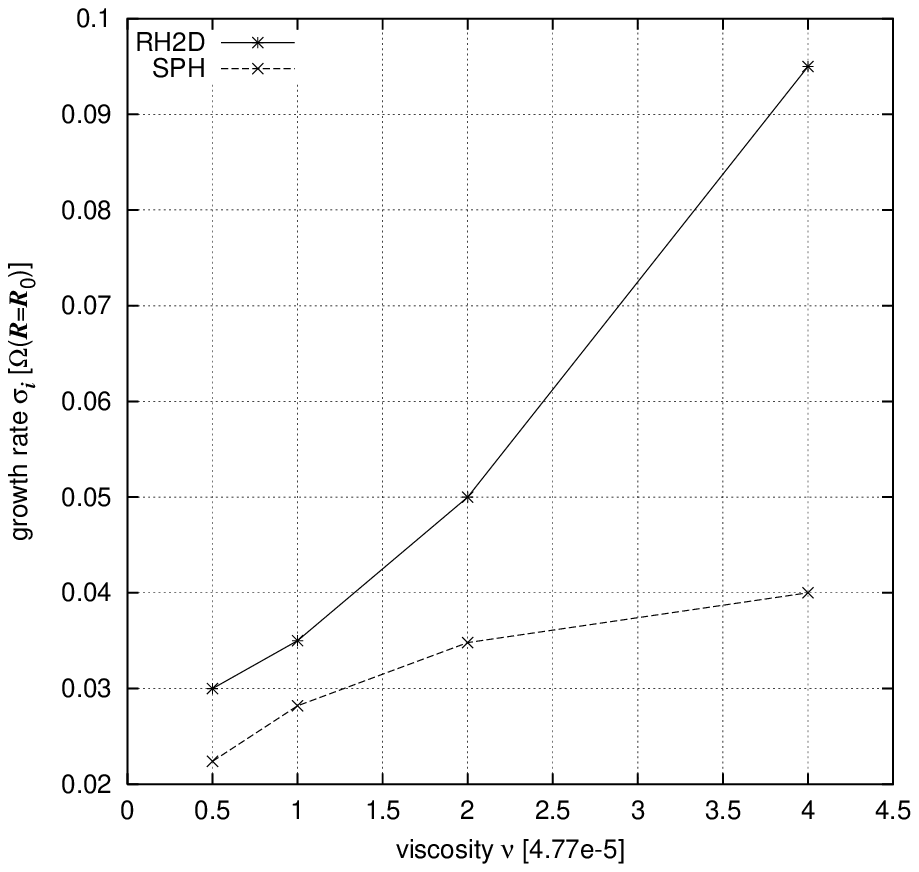}}
\end{center}
  \caption{The growth rates (in units of $\Omega$ at $R=1$)
   versus kinematic viscosity in units
   of the standard viscosity ($4.77 \times 10^{-5}$)
   for different kinematic viscosities
   and for both numerical methods.
   \label{fig:growth}
    }
\end{figure}
\subsubsection{Varying resolution and viscosity}
To test the influence of the numerical resolution we first fixed
the number of grid points in the radial direction (to 128) and then
varied the number of the angular grid points (from 128 to 370).
The results in Fig.~\ref{fig:f128} indicate that the growth rate
of the $m=1$ mode (measured at $R=1 R_0$) is nearly independent
of the resolution in $\varphi$
and converges for large $N_\varphi$ to one value.
This is consistent with our result that to second order the growth rates
should be independent of the azimuthal wave number $m$.

In the next set of models the number of radial grid points was
allowed to vary as well.
The time evolution of the total radial kinetic energy is
displayed in Fig.~\ref{fig:ekin_r} for models with different
radial (and azimuthal) grid resolutions.
For very small radial resolutions $N_R=64$ there is very little or no
growth. Then, for an increasing number of radial grid points
the growth rates increase as well.
This effect of faster growth for higher resolution is not an
artifact of the numerical analysis but is to be expected from
our analytical estimates. It was shown that the growth
rates $\sigma$ depend on the radial wave number $k$:
The smaller the wavelength of the perturbation the faster the
growth rates. But the smallest wavelength that can be resolved by numerical
computations depends naturally on the resolution.
Notice, that the wave crests in such a simulation are always resolved
by only a few grid cells (see Fig.~\ref{fig:ring6.ave}).

To study the influence of the kinematic viscosity we varied $\nu$
from the standard value to higher and smaller values. 
In Fig.~\ref{fig:visc} results are shown for several models.
A measure of the global deviation $\Delta \Sigma$ of the density
from the axisymmetric structure has been plotted, with
$\Delta \Sigma$ defined as
\begin{equation}
       \Delta \Sigma = \max_{\cal D} \left( 
     \frac{\Sigma (r, \,  \varphi) - \Sigma (r, \, \varphi + \pi) } 
          {\Sigma (r, \, \varphi) + \Sigma (r, \, \varphi + \pi) } \right) 
\label{eq:deltasigma}
\end{equation}
where the maximum is taken over the whole computational domain ${\cal D}$.
This measurement of $\Delta \Sigma$ is often a better tool to analyse
the growth rates than just studying the $m=1$ mode at some specific radius
as done above.

It can be seen that upon increasing the viscosity the
growth rates are also increasing, which is again in agreement
with our analytical estimates.
The initial increase of  $\Delta \Sigma$ for the two models
with the lowest viscosities ($\nu_s$ and $1/2\; \nu_s$) is related to the
nearly ring like disturbances described before. These decline
first, and then later the growth of the spiral disturbance sets
in. The exact behaviour of the curves depends on the initial random
perturbation as well.

Finally we display the obtained growth rates for different viscosity
coefficients in Fig.~\ref{fig:growth}, where the solid line denotes
the results computed by the finite difference method. 
For larger viscosities the growth rates
are also larger which is in agreement with Eq.~(\ref{grst}). 
To compare exactly with the analytical results, the
wavelength of the most unstable mode needs to be known as well.
However, the wave number is a function of radius, making a detailed comparison
not as easy. It is found in general that the obtained 
growth rates (as given for example in Fig.~\ref{fig:growth}) are in the
right order with the first term of Eq.~(\ref{grst}) but are
consistently lower.
The radial variation of the quantities may easily account for this
slight discrepancy.

Additionally, results of similar SPH calculations are plotted in
Fig.~\ref{fig:growth}, indicated by the dashed line. In all
cases, the SPH results fit the dispersion relation (\ref{grst}) well
within the accuracy of the SPH simulations. However, the growth rates
as well as the associated wave numbers of the most unstable mode are
lower than the grid-based results. Moreover, the growth rate increases
slower with increasing viscosity. 
This behaviour may demonstrate that the relation of growth rate and
wave number is depending on spatial resolution, initial condition, and
on the numerical method.
The discrepancy may also be caused by 
different evaluation methods. 
While for the SPH results the growth rates were measured locally from
the evolution of the amplitude of the $m=1$ Fourier mode at radius $R
= R_0$, for the grid-based results they were determined globally by
analysing the change of $\Delta \Sigma$ according to
Eq.~(\ref{eq:deltasigma}). 
This may also be another cause why the grid-based results fulfil the
dispersion relation (\ref{grst}) not exactly.
\section{Conclusion}
\label{sec:conclusion}
We have shown that the spiral instability found in various simulations
of the viscously evolving dust ring can be understood in terms
of a secular spiral instability driven
by viscosity. To accomplish this, we performed
a perturbation analysis using a time-stretching transformation. It
turned out that the widely known analytic solutions for the
surface density evolution and the azimuthal velocity of the viscously
spreading ring consist of the zeroth order terms 
of the expansion, while 
the solution for the radial velocity is given by the 1st order terms.

As a result of the perturbation analysis we found that the ring may
develop one-armed spiral structures to 1st order and one- and two-armed
spiral features to 2nd order. A local stability analysis of an
eccentricity function $E$ provided a dispersion relation that led 
for a constant kinematic viscosity to a
growth rate for secular spiral instabilities of 
\begin{displaymath}
   \sigma_i = \frac{1}{3} \, k^2 \, \nu \, + \, \mathcal{O}(\nu/R^2)
\end{displaymath}
in the limit $kR \gg 1$.
The exact form of the growth rate depends on the viscosity prescription
$\nu = \nu (R)$.
Our theoretical results could be confirmed in several
simulations using two different numerical methods.

As a consequence for numerical simulations of accretion discs, the
spiral instability has to be taken into account when using 
the spreading ring for instance as test problem for code development.
Otherwise, occasionally emerging spiral features may be mistaken
for numerical instabilities of the algorithms used for the calculation of the
viscous forces.

The physical implications of our results are not as obvious. Because
the kinematic viscosity $\nu$ is held axisymmetric throughout the
whole stability analysis, the discovered non-axisymmetric
instabilities may be of importance mainly in accretion discs where the
viscosity is determined uniformly by external effects like
irradiation.

\begin{acknowledgements}
We would like to thank the late Harald Riffert
who contributed in many fruitful
discussions substantially to this work and who suggested the use of
the stretching transformation for the perturbation analysis.
We also want to thank the referee, G.~Ogilvie, for many
helpful suggestions to improve and clarify this paper.

Research in theoretical astrophysics at the University of Leicester is
supported by a PPARC rolling grant. RS gratefully acknowledges funding
by this PPARC rolling grant. 

RS wishes to acknowledge the support of Advanced Micro Devices (AMD)
for the Leicester Theoretical Astrophysics Group's Linux cluster on
which some of the calculations were performed. Parts of his
simulations were also performed on the GRAND supercomputer funded by a
PPARC HPC grant.
\end{acknowledgements}

\bibliographystyle{aa}
\bibliography{h3873}

\end{document}